\documentclass[aps,pre,twocolumn,superscriptaddress,showpacs]{revtex4}
\usepackage{amssymb,amsmath,latexsym}
\usepackage{epsfig}
\usepackage{graphicx}
\usepackage{color}

\begin{document}

\title{Detrended fluctuation analysis made flexible to detect range of cross-correlated fluctuations}

\author{Jaros{\l}aw Kwapie\'n}
\author{Pawe{\l} O\'swi\c ecimka}
\affiliation{Institute of Nuclear Physics, Polish Academy of Sciences, Krak\'ow, Poland}
\author{Stanis{\l}aw Dro\.zd\.z}
\affiliation{Institute of Nuclear Physics, Polish Academy of Sciences, Krak\'ow, Poland}
\affiliation{Faculty of Physics, Mathematics and Computer Science, Cracow University of Technology, Krak\'ow, Poland}

\date{\today}

\begin{abstract}

The detrended cross-correlation coefficient $\rho_{\rm DCCA}$ has recently been proposed to quantify the strength of cross-correlations on different temporal scales in bivariate, non-stationary time series. It is based on the detrended cross-correlation and detrended fluctuation analyses (DCCA and DFA, respectively) and can be viewed as an analogue of the Pearson coefficient in the case of the fluctuation analysis. The coefficient $\rho_{\rm DCCA}$ works well in many practical situations but by construction its applicability is limited to detection of whether two signals are generally cross-correlated, without possibility to obtain information on the amplitude of fluctuations that are responsible for those cross-correlations. In order to introduce some related flexibility, here we propose an extension of $\rho_{\rm DCCA}$ that exploits the multifractal versions of DFA and DCCA: MFDFA and MFCCA, respectively. The resulting new coefficient $\rho_q$ not only is able to quantify the strength of correlations, but also it allows one to identify the range of detrended fluctuation amplitudes that are correlated in two signals under study. We show how the coefficient $\rho_q$ works in practical situations by applying it to stochastic time series representing processes with long memory: autoregressive and multiplicative ones. Such processes are often used to model signals recorded from complex systems and complex physical phenomena like turbulence, so we are convinced that this new measure can successfully be applied in time series analysis. In particular, we present an example of such application to highly complex empirical data from financial markets. The present formulation can straightforwardly be extended to multivariate data in terms of the $q$-dependent counterpart of the correlation matrices and then to the network representation.

\end{abstract}


\maketitle

\section{Introduction}
\label{sect.1}

Standard correlation measures like the Pearson correlation coefficient and the cross-correlation function require stationary data in order to provide reliable results, which is a requirement that is hard to fulfill in many real-world situations (the financial and physiological data are the negative examples here~\cite{pagan2002,mikosch2004,bassler2007,schmitt2013,mayer-kress1994,peng1995,kohlmorgen2000,bernaola-galvan2001,perello2008,kwapien2014}). (By stationarity we mean stability of the probability distribution functions of the data over time; from this perspective nonstationarity can be produced both by the long-range autocorrelations and by the pdf's heavy tails that make any signal length effectively insufficient.) This problem can to some degree be resolved by replacing original signals with the corresponding detrended fluctuations, i.e. by considering the power-law correlations between the polynomially detrended walks, which are much more stationary than the orginal data. This approach was introduced in a context of autocorrelations in~\cite{peng1994} as the detrended fluctuation analysis (DFA) and immediately gained popularity among the researchers working with empirical data across many disciplines. The modification of DFA oriented towards detection of the power-law cross-correlations is known as the detrended cross-correlation analysis (DCCA)~\cite{podobnik2008} and it can be used in analysis of bivariate and multivariate empirical data~\cite{podobnik2009,xu2010,zebende2011a,qian2015}. 

Among the reasons behind the popularity of DFA was its ability of detecting fractal character of signals, which was subsequently extended to the multifractal case (the MFDFA method~\cite{kantelhardt2002}), which also proved very useful if applied to empirical data~\cite{oswiecimka2005,muzy2008,ivanov1999,udovichenko2002,witt2013,calvet2002,drozdz2009,drozdz2010,koscielny2006,%
kantelhardt2006,ausloos2012,jafari2007,oswiecimka2011,grech2008,maiorino2015,makowiec2009,kristoufek2015a,drozdz2015,rotundo2015}, especially owing to its superior reliability if compared to other methods~\cite{oswiecimka2006}. DCCA was also generalized in order to be applicable to signals with multifractal cross-correlations and the resulting MFDCCA/MFDXA algorithm~\cite{zhou2008} also attracted some attention~\cite{jiang2011,he2011,li2012,wang2012}. However, this generalization raises controversy as at some stage it requires to neglect the detrended covariance sign to avoid obtaining complex values, which leads to inevitable loss of information about analyzed signals and, consequently, to incorrect results~\cite{oswiecimka2014}. This drawback has recently been removed with a new sign-sensitive method of the multifractal detrended cross-correlation analysis with an acronym MFCCA~\cite{oswiecimka2014,chen-hua2015} that is more consistent than MFDCCA as a multifractal generalization of DCCA (for more details see Section II).

Although the DFA and DCCA methods were designed to deal with nonstationary signals, they are related to each other exactly in the same way as the variance and the covariance analyses are related to each other in the case of stationary data. Therefore, by exploiting both these methods, an analogue of the Pearson coefficient was introduced. It is known as the detrended cross-correlation coefficient $\rho_{\rm DCCA}$~\cite{zebende2011b} and serves as a tool for quantifying strength of correlations in the fluctuations of detrended signals at a given time scale~\cite{zebende2011b,vassoler2012,zebende2013,reboredo2014}. The main advantage of $\rho_{\rm DCCA}$ over the Pearson coefficient is its ability to quantify correlations in nonstationary signals~\cite{kristoufek2014a}. What is important is that the signals under study may either be fractal or non-fractal since $\rho_{\rm DCCA}$ is defined for a single scale. It is also worth mentioning that there is a counterpart of $\rho_{\rm DCCA}$ constructed for the detrended moving average cross-correlation analysis (DMCA)~\cite{kristoufek2014b} but considering it here exceeds the objectives of this work.

By definition, the detrended cross-correlation coefficient is insensitive to higher-order statistics of fluctuations, i.e., other than the simple covariance. This means that values of $\rho_{\rm DCCA}$ cannot indicate whether the detected cross-correlations between two signals originate from the fluctuations of all amplitudes equally likely or rather some specific range of amplitudes gives a dominant contribution, while the remaining fluctuations can be much less correlated or even completely uncorrelated. One might easily imagine situations, in which such insensitivity can be considered a serious disadvantage of the method. For instance, let each nonstationary signal in a pair be a mixture of two processes, one of which is correlated in both signals and has a relatively small amplitude, while the other is unique to each signal and has a relatively high amplitude. The coefficient $\rho_{\rm DCCA}$ would probably indicate that the signals are somehow cross-correlated but it would not bring any information allowing one to identify the amplitude of the cross-correlated components.

In order to avoid such insensitivity, here we propose a generalization of the detrended cross-correlation coefficient in such a way that it becomes sensitive to correlations in the fluctuations of a selected amplitude range. The easiest way to do this is by introducing a $q$-dependent detrended cross-correlation coefficient $\rho_q$ ($q \in \mathcal{R}$) based on the so-called $q$-dependent fluctuation functions $F_q$ from MFDFA and MFCCA~\cite{kantelhardt2002,oswiecimka2014}. The idea is based on the fact that from a number of different values contibuting to a sum, one can select specific values (e.g., the large, the medium, or the small ones) by rising all the values in the sum to some power (high positive, small positive, or negative one, respectively). Like $\rho_{\rm DCCA}$, the coefficient $\rho_q$ is not related to fractal properties of signals so it can be used to quantify cross-correlations between any signals.

As by its definition $\rho_q$ is intended to be a tool for analyzing nonstationary signals, we expect that it can find broad applications in the studies of empirical data from natural complex systems: physical, biological, social, financial, etc. In the remaining part of the paper, we present the formal definition of $\rho_q$ (Section II), show examples of how it works if applied to the computer-generated signals representing different stochastic processes (Section III) and to empirical data from the financial markets (Section IV), and finally present the main conclusions (Section V).

\section{The $q$-dependent detrended cross-correlation coefficient}
\label{sect.2}

A fundamental quantity for the detrended fluctuation analysis as well as for any type of its derivative methods is the variance (covariance) $f_{ZZ}^2$ ($f_{XY}^2$) of the detrended signals $X,Y$ ($Z$ stands for either $X$ or $Y$). Let us consider a pair of time series ${x(i)}_{i=1,...,T}$ and ${y(i)}_{i=1,...,T}$ divided into $2 M_s$ separate boxes of length $s$ (i.e., $M_s$ boxes starting from the opposite ends). A detrending procedure consists of calculating in each box $\nu$ ($\nu=0,...,2 M_s - 1$) the residual signals $X,Y$ equal to the difference between the integrated signals and the $m$th-order polynomials $P^{(m)}$ fitted to these signals:
\begin{eqnarray}
X_{\nu}(s,i) = \sum_{j=1}^i x(\nu s + j) - P_{X,s,\nu}^{(m)}(j),\\
Y_{\nu}(s,i) = \sum_{j=1}^i y(\nu s + j) - P_{Y,s,\nu}^{(m)}(j).
\end{eqnarray}
In the present work we use $m=2$. The covariance and the variances of $X$ and $Y$ in a box $\nu$ are defined as:
\begin{eqnarray}
\label{eq::covariance}
f_{XY}^2(s,\nu) = {1 \over s} \sum_{i=1}^s X_{\nu}(s,i) Y_{\nu}(s,i),\\
f_{ZZ}^2(s,\nu) = {1 \over s} \sum_{i=1}^s Z_{\nu}^2(s,i),
\label{eq::variance}
\end{eqnarray}
where $Z$ again means either $X$ or $Y$. These quantities can be used to define a family of the so-called fluctuation functions of order $q$~\cite{kantelhardt2002,oswiecimka2014}:
\begin{eqnarray}
\label{eq::covariance.q}
F_{XY}^q(s) = {1 \over 2 M_s} \sum_{\nu=0}^{2 M_s - 1} {\rm sign} \left[f_{XY}^2(s,\nu)\right] |f_{XY}^2(s,\nu)|^{q/2},\\
F_{ZZ}^q(s) = {1 \over 2 M_s} \sum_{\nu=0}^{2 M_s - 1} \left[f_{ZZ}^2(s,\nu)\right]^{q/2}.
\label{eq::variance.q}
\end{eqnarray}
The above definition of $F_{XY}^q(s)$ guarantees that: (i) no imaginary part occurs in $F_{XY}^q(s)$ (only the absolute values are raised to a real power $q/2$) and (ii) by preserving signs of the covariances $f_{XY}^2(s,\nu)$, no information is lost while taking the absolute values. For $q=2$ the above definitions are reduced to a simpler form:
\begin{eqnarray}
\label{eq::covariance.average}
F_{XY}^2(s) = {1 \over 2 M_s} \sum_{\nu=0}^{2 M_s - 1} f_{XY}^2(s,\nu),\\
F_{ZZ}^2(s) = {1 \over 2 M_s} \sum_{\nu=0}^{2 M_s - 1} f_{ZZ}^2(s,\nu)
\label{eq::variance.average}
\end{eqnarray}
that may be interpreted as an average covariance and average variances for the boxes of size $s$. In the standard application of the fluctuation functions in MFDFA and MFCCA, one observes the dependence of $F_{XY}^q$ and $F_{ZZ}^q$ on the scale $s$ and looks for a convincing scaling behavior: $\left[ F_{XY}^q(s) \right]^{1/q} \sim s^{\gamma(q)}$ and/or $\left[ F_{ZZ}^q(s) \right]^{1/q} \sim s^{\delta(q)}$, which indicates a fractal structure of the signals (monofractal for the constant functions: $\gamma(q)= c$ and $\delta(q)= c$, multifractal otherwise).

The structure of Eqs.~(\ref{eq::covariance}) and (\ref{eq::variance}), which resembles the ordinary covariance and variance, respectively, suggested one to introduce the detrended cross-correlation coefficient~\cite{zebende2011b}:
\begin{equation}
\rho_{\rm DCCA}(s) = {F_{XY}^2(s) \over \sqrt{ F_{XX}^2(s) F_{YY}^2(s) }}.
\label{eq::rho.dcca}
\end{equation}
Owing to its normalized range of values~\cite{podobnik2011}: $-1 \le \rho_{\rm DCCA} \le 1$ with $\rho_{\rm DCCA} = 0$ in the case of uncorrelated signals, $\rho_{\rm DCCA}=1$ in the case of a perfect cross-correlation, and $\rho_{\rm DCCA} = -1$ in the case of a perfect anticorrelation, the coefficient $\rho_{\rm DCCA}(s)$ can be used to quantify the strength of cross-correlations between the detrended signals $X,Y$ on different scales $s$ and to compare this strength among different signal pairs~\cite{zebende2011b}.

The structure of Eqs.~(\ref{eq::covariance.average}) and (\ref{eq::variance.average}) indicates that all the boxes contribute to the fluctuation functions and the correlation coefficient with the same weight, irrespective of how large (or how small) is $f_{XY}^2(s,\nu)$ and $f_{ZZ}^2(s,\nu)$ in a particular box $\nu$. This means that, by using solely the coefficient $\rho_{\rm DCCA}$, it is impossible to observe how the boxes that are characterized by the fluctuations of a specific amplitude range contribute to the overall cross-correlations. The following definition of a new $q$-dependent detrended cross-correlation ($q$DCCA) coefficient $\rho_q(s)$ allows one to overcome this constraint:
\begin{equation}
\rho_q(s) = {F_{XY}^q(s) \over \sqrt{ F_{XX}^q(s) F_{YY}^q(s) }}
\label{eq::rho.q}
\end{equation}
as the real exponent $q$ plays the role of a respective filter here. For $q=2$ we restore the definition (\ref{eq::rho.dcca}) of $\rho_{\rm DCCA}$, for $q > 2$ the boxes with high values of $f_{XY}^2(s,\nu)$ and $f_{ZZ}^2(s,\nu)$ contribute to $\rho_q(s)$ the most, while for $q < 2$ the boxes with relatively small values do it the most significantly. The more deviated from the value $q=2$ the exponent $q$ is, the more extreme fluctuations in the corresponding boxes contribute to the coefficient $\rho_q(s)$. Note that no form of the $s$-dependence in Eqs.~(\ref{eq::covariance.q}) and~(\ref{eq::variance.q}) has been assumed, so the signals under study do not need to be fractal at all.

For $q \ge 0$, values of the $\rho_q$ coefficient are bound within the same range as $\rho_{\rm DCCA}$, i.e.,
\begin{equation}
-1 \le \rho_q \le 1 .
\label{eq::normalized}
\end{equation}
In order to show this, we fix the scale $s$ and prove the following Cauchy-Schwarz-like inequality:
\begin{equation}
\left[ F_{XY}^q(s) \right]^2 \le F_{XX}^q(s) F_{YY}^q(s), \quad q \ge 0.
\label{eq::cauchy-schwarz-like}
\end{equation}
First, observe that from the relation: $a^{2\alpha}+b^{2\alpha} \ge 2 a^{\alpha} b^{\alpha}$ ($a,b \ge 0$, $\alpha > 0$) it follows for any two boxes $\nu,\mu$ that:
\begin{eqnarray}
\nonumber
2 \left[ f_{XX}^2(s,\nu) f_{YY}^2(s,\nu) f_{XX}^2(s,\mu) f_{YY}^2(s,\mu) \right]^{\alpha} \le \\
\nonumber
\le  \left[ f_{XX}^2(s,\nu) f_{YY}^2(s,\mu) \right]^{2 \alpha} + \left[ f_{XX}^2(s,\mu) f_{YY}^2(s,\nu) \right]^{2 \alpha} \\
\quad
\label{eq::lemma}
\end{eqnarray}
since $f_{ZZ}^2(s,\nu) \ge 0$. Now, to simplify the notation, we temporarily neglect the signum function in Eq.~(\ref{eq::covariance.q}) and assume that all the covariances $f_{XY}^2(s,\nu)$ are positive. We then start from the l.h.s. of Eq.~(\ref{eq::cauchy-schwarz-like}):
\begin{eqnarray}
\nonumber
\left[ F_{XY}^q(s) \right]^2 = {1 \over 4 M_s^2} \lbrace \sum_{\nu=0}^{2 M_s-1} \left[ f_{XY}^2(s,\nu) \right]^{q/2} \rbrace^2  = \\
\nonumber
= {1 \over 4 M_s^2} \sum_{\nu=0}^{2 M_s-1} \left[ f_{XY}^2(s,\nu) \right]^q + \\
\nonumber
+ {1 \over 4 M_s^2} \sum_{\nu=0}^{2 M_s-1} \sum_{\mu=\nu+1}^{2 M_s-1} 2 \left[ f_{XY}^2(s,\nu) f_{XY}^2(s,\mu) \right]^{q/2} \le \\
\nonumber
\le {1 \over 4 M_s^2} \sum_{\nu=0}^{2 M_s-1} \left[ f_{XX}^2(s,\nu) f_{YY}^2(s,\nu) \right]^{q/2} + \\
\nonumber
+{1 \over 4 M_s^2} \sum_{\nu=0}^{2 M_s-1} \sum_{\mu=\nu+1}^{2 M_s-1} \lbrace \left[ f_{XX}^2(s,\nu) f_{YY}^2(s,\mu) \right]^{q/2} + \\
\nonumber
+ \left[ f_{XX}^2(s,\mu) f_{YY}^2(s,\nu) \right]^{q/2} \rbrace = \\
\nonumber
= {1 \over 4 M_s^2} \sum_{\nu=0}^{2 M_s-1} \sum_{\mu=0}^{2 M_s-1} \left[ f_{XX}^2(s,\nu) f_{YY}^2(s,\mu) \right]^{q/2} = \\
\nonumber
= F_{XX}^q(s) F_{YY}^q(s) ,
\end{eqnarray}
where we exploited the Cauchy-Schwarz inequality: $\left[ f_{XY}^2(s,\nu) \right]^2 \le f_{XX}^2(s,\nu) f_{YY}^2(s,\nu)$, the implication: $|a| \le |b| \Rightarrow |a|^q \le |b|^q$ for $q \ge 0$, and Eq.~(\ref{eq::lemma}).

In general, the covariances $f_{XY}^2(s,\nu)$ can be negative and we must take this fact into consideration. Fortunately, the following relation always holds:
\begin{eqnarray}
\nonumber
- \sqrt { F_{XX}^q(s) F_{YY}^q(s) } \le - {1 \over 2 M_s} \sum_{\nu=0}^{2 M_s - 1} |f_{XY}^2(s,\nu)|^{q/2} \le \\
\nonumber
\le {1 \over 2 M_s} \sum_{\nu=0}^{2 M_s - 1} {\rm sign} \left[f_{XY}^2(s,\nu)\right] |f_{XY}^2(s,\nu)|^{q/2} \le \\
\nonumber
\le {1 \over 2 M_s} \sum_{\nu=0}^{2 M_s - 1} |f_{XY}^2(s,\nu)|^{q/2} \le \sqrt { F_{XX}^q(s) F_{YY}^q(s) },
\end{eqnarray}
which ends the proof of the inequalities~(\ref{eq::cauchy-schwarz-like}) and~(\ref{eq::normalized}).

For $q < 0$ the situation looks different, because the implication $|a| \le |b| \Rightarrow |a|^q \le |b|^q$ is false in this case. This means that the denominator in Eq.~(\ref{eq::rho.q}) may be arbitrarily small as compared to the numerator modulus and $\rho_q(s)$ can then assume either large positive or large negative values: $|\rho_q(s)| \gg 1$ for some scales $s$. This can be evident for uncorrelated or partially correlated signals, while it is unlike for perfectly correlated ones. Therefore, the interpretation of the values of $\rho_q(s)$ for $q < 0$ is a delicate issue that will be discussed in more detail in Section III.

\section{Computer-generated signals}
\label{sect.3}

In this Section we present examples of the application of the new coefficient $\rho_q$ to the computer-generated time series representing different stochastic processes, in which the cross-correlations can fully be controlled, and then we discuss the performance of the coefficient in those cases. Since $\rho_q$ is oriented towards analysis of nonstationary data, we prefer to employ stochastic models that produce signals with long memory and/or heavy-tailed probability distribution functions. Long memory, which produces trends, and heavy tails of the fluctuations' p.d.f.s, which can produce apparent trends, are the principal sources of nonstationarity in data from natural complex systems, therefore their presence in the model signals is highly desired.

We focus on two models: the autoregressive fractionally integrated moving average (ARFIMA) and the Markov-switching multifractal (MSM). The former produces fractal signals with long memory and, despite its origin in financial economics~\cite{hosking1981}, it is broadly used to model anomalous diffusion in various fields of science, like atmosphere physics and geophysics~\cite{gil-alana2012,yusof2015}, astrophysics~\cite{burnecki2014}, biology and physiology~\cite{torre2007,leite2013}, and many other. The latter can be viewed as a version of a random walk in random time~\cite{saakian2012}. It is based on the stochastic multiplicative cascades and creates signals with long memory, heavy-tailed p.d.f., and with arbitrary length (unlike more typical cascading processes whose length is limited to the consecutive powers of the number of branches)~\cite{lux2006}. MSM can be exploited to model various cascade-like phenomena, e.g., turbulence~\cite{rypdal2011} and financial volatility~\cite{lux2006,calvet2008} and to predict future evolution of the corresponding observables.

\subsection{Perfectly cross-correlated signals}

\begin{figure}
\includegraphics[scale=0.32]{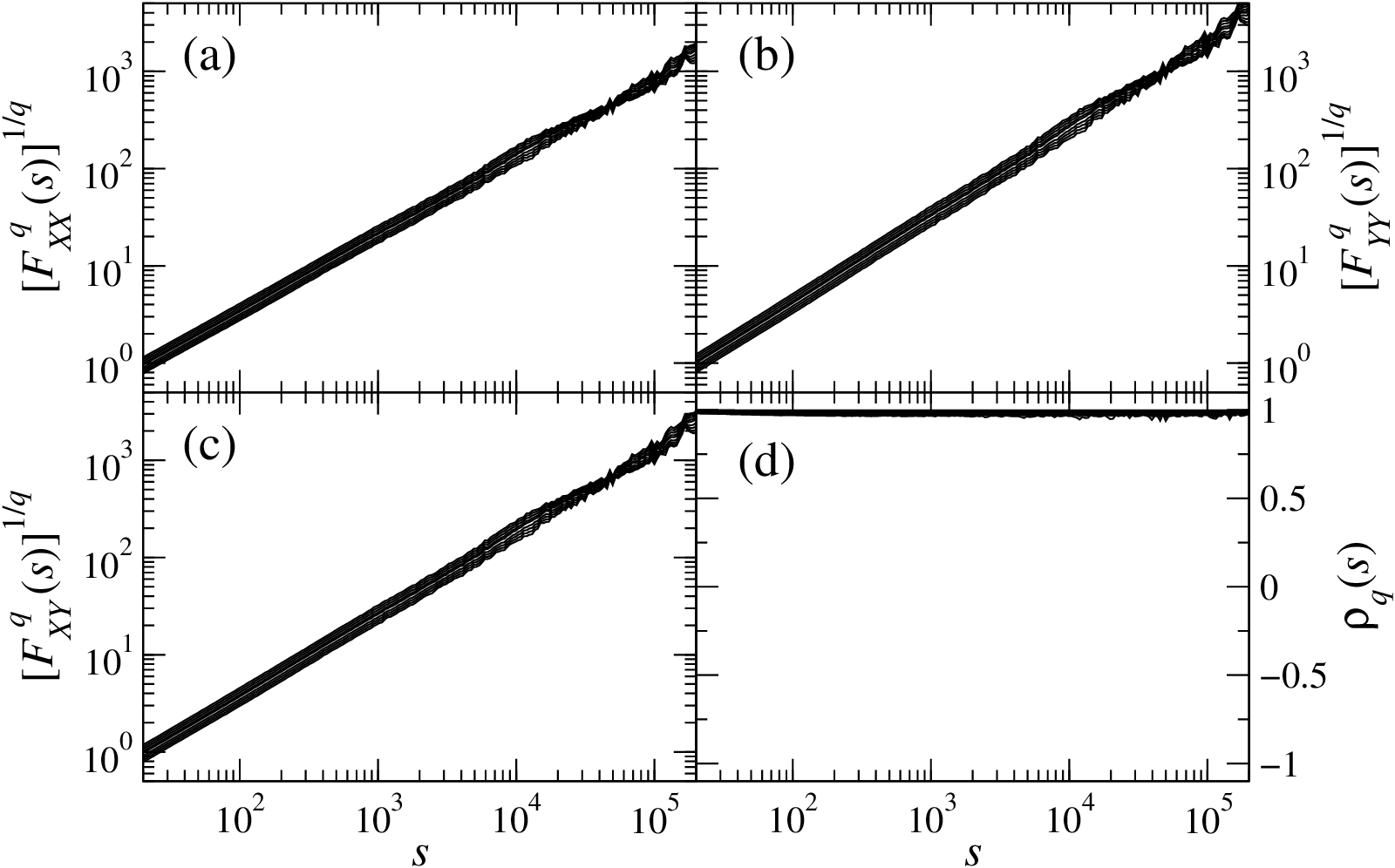}
\caption{(Color online) (a)-(b) The $q$-dependent fluctuation functions $\left[ F_{XX}^q(s) \right]^{1/q}$ and $\left[ F_{YY}^q(s) \right]^{1/q}$ for two ARFIMA time series sharing the same Gaussian noise term. A family of the fluctuation functions for different values of $q$, $q \in \langle -4,4 \rangle$ are shown for each time series (the topmost one represents $q=4$). The linear dependence of $\left[ F_{XX}^q(s) \right]^{1/q}$ and $\left[ F_{YY}^q(s) \right]^{1/q}$ on the double logarithmic plots indicates a fractal character of the analyzed signals. (c) The $q$-dependent cross-fluctuation function $\left[ F_{XY}^q(s) \right]^{1/q}$ for the same pair of the time series indicating fractal cross-correlations between them. (d) The $q$-dependent detrended cross-correlation coefficient $\rho_q(s)$ for different values of $q$. Note the almost perfect cross-correlation for any $s$ and $q$.}
\label{fig::arfima.correlated}
\end{figure}

First, we study behavior of $\rho_q(s)$ for a pair of maximally correlated signals. In order to prepare such signals, we use the ARFIMA process~\cite{hosking1981} defined by the following formulas:
\begin{eqnarray}
x(i) = \sum_{j=1}^{\infty} a_j(d_x)x(i-j) + \varepsilon(i) \\
\nonumber
y(i) = \sum_{j=1}^{\infty} a_j(d_y)y(i-j) + \varepsilon(i),
\label{eq::arfima}
\end{eqnarray}
where the parameters $d_z$ ($z \equiv x,y$), fulfilling the condition: $-1/2 < d_z < 1/2$, characterize the temporal range of the linear autocorrelations in $x(i), y(i)$ and are strictly related to the Hurst exponents: $H=1/2+d_z$. The quantity $a_j(d_z)$ is called weight and defined by:
\begin{equation}
a_j(d_z)= {\Gamma(j-d_z) \over \Gamma(-d_z) \Gamma(1+j)}.
\end{equation}
The time series $x(i)$ and $y(i)$ are correlated due to a common noise term $\varepsilon(i)$ being an i.i.d.~Gaussian random variable. The ARFIMA signals are (mono)fractal, so their fluctuation functions $\left[ F_{ZZ}^q(s) \right]^{1/q}$ form the families of power-laws presented in the upper panels of Fig.~\ref{fig::arfima.correlated}. The fractally cross-correlated nature of $x(i)$ and $y(i)$ can be seen in Fig.~\ref{fig::arfima.correlated}(c) as a family of parallel power-law functions $\left[ F_{XY}^q(s) \right]^{1/q}$. Careful inspection of the three panels: Fig.~\ref{fig::arfima.correlated}(a)-(c) suggests that the fluctuation amplitudes of the signals under study are maximally cross-correlated, so we may expect that a properly defined $q$DCCA coefficient should reflect this fact with its value being close to unity. Indeed, in Fig.~\ref{fig::arfima.correlated}(d), $\rho_q(s) \simeq 1$ irrespective of $s$ and $q$. This invariance can also be a consequence of the stationarity of the detrended ARFIMA processes. For signals with heavy tails of the pdfs (which preserve their non-stationary character after detrending), we would expect more sizable deviations of $\rho_q(s)$ from unity even if such signals were on average strongly cross-correlated. This is because large fluctuations that may occur in different signals at different moments can have strong impact on covariance of the signals in the related boxes (up to the considerably large widths $s$), which can suppress it in these particular boxes and thus influence $F_{XY}^q(s)$. The higher $|q|$ is, the stronger effect of this type may happen.

\subsection{Uncorrelated signals}

\begin{figure}
\includegraphics[scale=0.32]{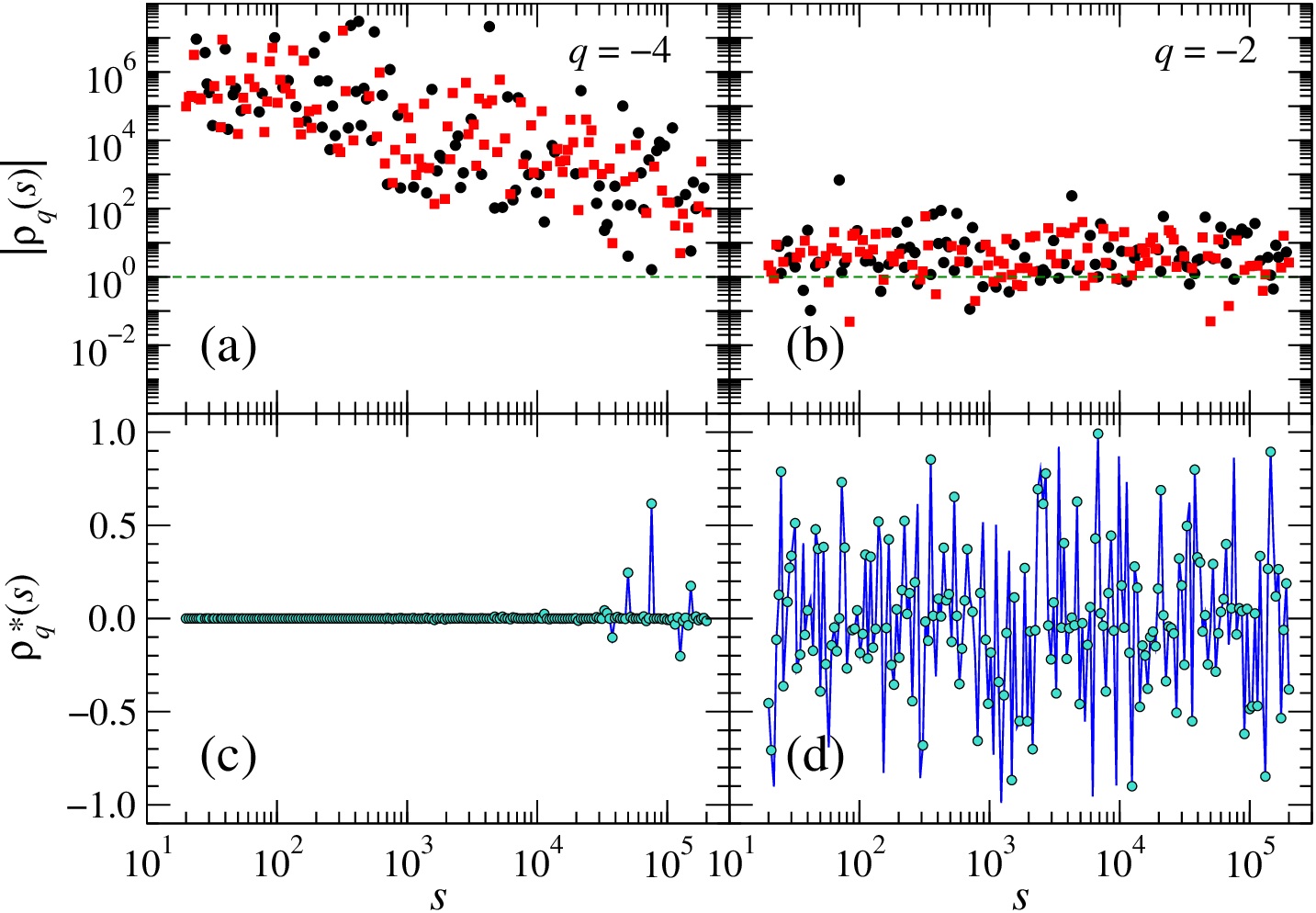}
\caption{(Color online) The $q$-dependent detrended cross-correlation coefficient $\rho_q$ as a function of the temporal scale $s$ for sample values of $q < 0$: $q=-4$ (left) and $q=-2$ (right). (Top) Scale dependence of the originally defined $\rho_q(s)$ for which values outside the standard interval $\langle -1,1 \rangle$ are allowed (the modulus $|\rho_q(s)|$ is taken in order to show both the positive (black circles) and the negative (red/grey squares) values on the logarithmic axis). Dashed lines indicate the $|\rho_q|=1$ level. (Bottom) The modified coefficient $\rho_q^{*}(s)$ defined by Eq.~(\ref{eq::rho.star}) with its values being in the interval $\langle -1,1 \rangle$. Those values of $\rho_q(s)$ that were inverted to obtain $\rho_q^{*}(s)$ are denoted by circles.}
\label{fig::arfima.unbound}
\end{figure}

In order to test the $q$DCCA coefficient $\rho_q(s)$ in the absence of cross-correlations, we take the same signals as before: $x(i)$ and $y(i)$ and destroy their temporal structure by random shuffling. However, before we show the complete results, we have to discuss the problem of $|\rho_q(s)| > 1$ for $q < 0$. Fig.~\ref{fig::arfima.unbound}(a)(b) shows $|\rho_q(s)|$ for (a) $q=-4$ and (b) $q=-2$. As one can see, its values can span a number of orders of magnitude with the more negative $q$ is, the larger this range can be. The unbound character of $\rho_q(s)$ makes any inference on the strength of the cross-correlations virtually impossible. As it has already been mentioned in Section II, the high values of $|\rho_q(s)|$ are produced if the denominator in Eq.~(\ref{eq::rho.q}) is much smaller than the numerator. This situation can predominantly happen if the signals under study are not cross-correlated or are cross-correlated weakly. This suggests that a value of $|\rho_q(s)|$ that deviates much from 1 in any direction may be viewed as an indicator of a lack of cross-correlations. Taking this into consideration, one can redefine the $q$DCCA coefficient in the following way:
\begin{equation}
\rho_q^{*}(s) =
\begin{cases}
\rho_q(s) & \textrm{if} \quad |\rho_q(s)| \le 1 \\
[\rho_q(s)]^{-1} & \textrm{if} \quad |\rho_q(s)| > 1.
\end{cases}
\label{eq::rho.star}
\end{equation}
Now the $\rho_q^{*}(s)$ coefficient remains always within the interval $\langle -1,1 \rangle$ even if $q < 0$. Two typical cases can be distinguished in this situation: (i) Fig.~\ref{fig::arfima.unbound}(c) exhibits that for $q=-4$ the coefficient $\rho_q^{*}(s)$ indicates the expected almost-zero level of cross-correlations; (ii) for $q=-2$ (Fig.~\ref{fig::arfima.unbound}(d)) $\rho_q^{*}(s)$ is clearly non-zero, but it is highly unstable with respect to $s$, bouncing between positive and negative values. Both these cases, if found in an analysis, may thus be considered the indicators of uncorrelated signals. For simplicity, from now on we will omit the star in the notation of the $q$DCCA coefficient:
\begin{equation}
\rho_q(s) \equiv \rho_q^{*}(s).
\end{equation}

\begin{figure}
\includegraphics[scale=0.48]{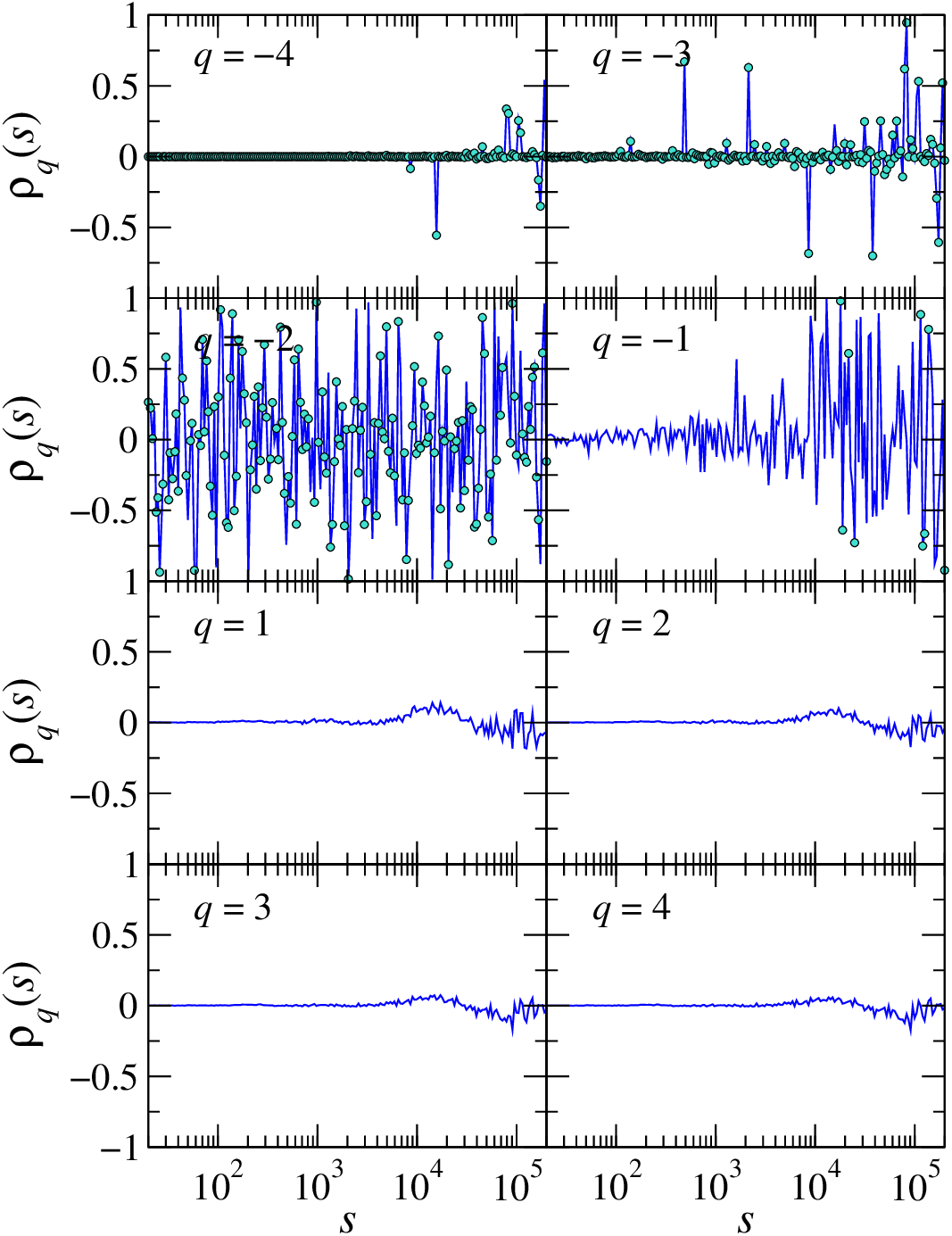}
\caption{(Color online) The coefficient $\rho_q(s)$ calculated for a pair of the randomized ARFIMA time series. In each panel, $\rho_q(s)$ for different $q$ is displayed. Statistical significance of the results can be assessed by comparing these results with Fig.~\ref{fig::arfima.deviation}. The inverted values of $\rho_q(s)$ for $q < 0$ are denoted by circles.}
\label{fig::arfima.shuffled}
\end{figure}

\begin{figure}
\includegraphics[scale=0.48]{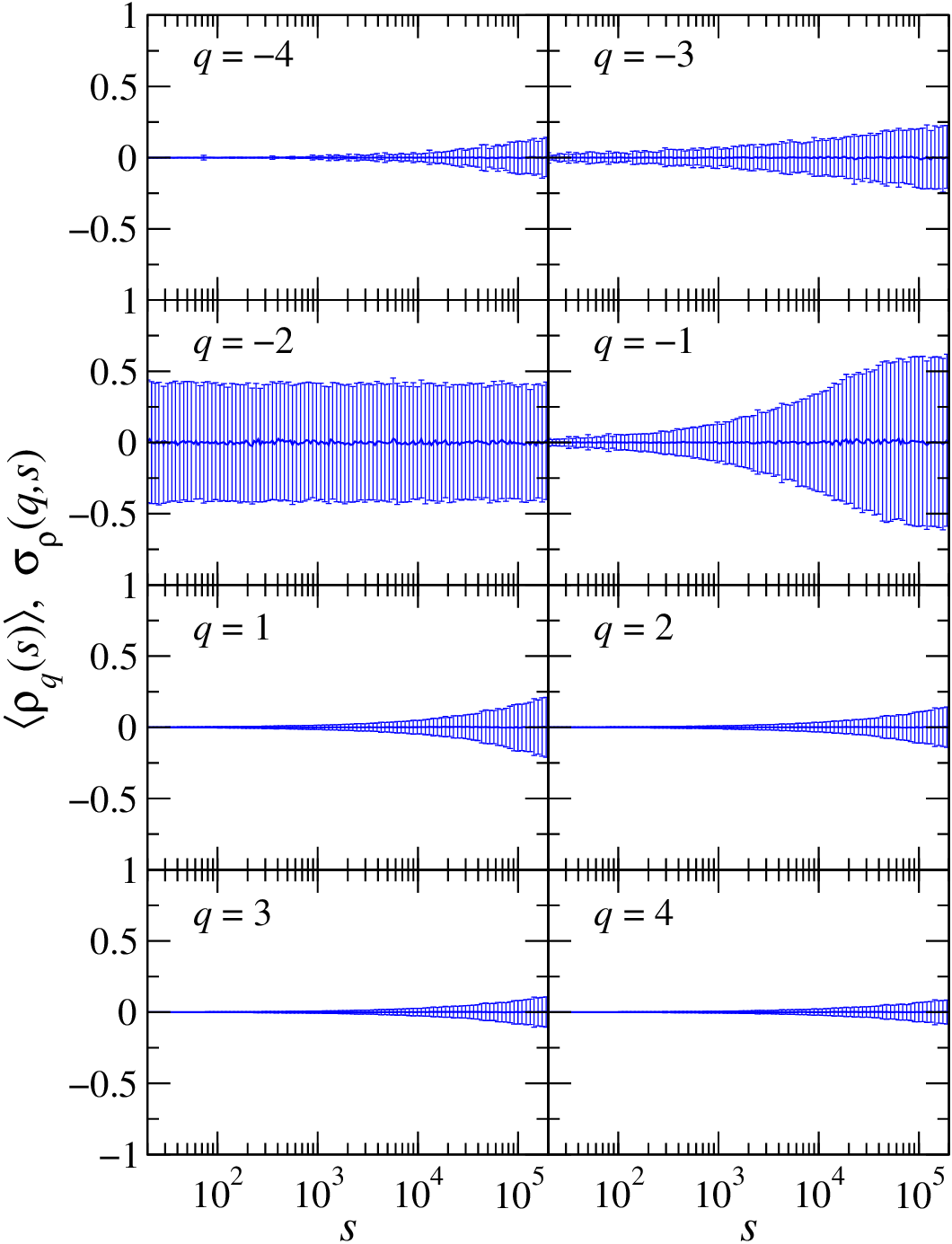}
\caption{(Color online) Mean $\langle \rho_q(s) \rangle$ (black solid lines) and standard deviation $\sigma_{\rho}(q,s)$ (blue/grey lines with error bars) of the coefficient $\rho_q(s)$ calculated for 10,000 pairs of the shuffled ARFIMA time series. In each panel, the result for different $q$ is displayed.}
\label{fig::arfima.deviation}
\end{figure}

Complete results of $\rho_q(s)$ for the randomized ARFIMA signals are presented in Fig.~\ref{fig::arfima.shuffled}. In the light of the above discussion, we are justified to conclude that the analyzed signals are not cross-correlated at any scale $s$ and at any $q<0$. Since for $q>0$ and $s>10^4$ we observe some deflections of $\rho_q(s)$ from zero, we have to perform a test that can resolve whether these deflections are statistically significant. In order to do this, we calculate $\rho_q(s)$ for $N=10,000$ pairs of the randomized ARFIMA signals and estimate the dispersion of $\rho_q(s)$. The related standard deviations $\sigma_{\rho}(q,s)$ are shown in Fig.~\ref{fig::arfima.deviation}. A direct comparison between Fig.~\ref{fig::arfima.shuffled} and Fig.~\ref{fig::arfima.deviation} leads us to a conclusion that the deflections from zero observed for $q>0$ in Fig.~\ref{fig::arfima.shuffled} cannot be considered statistically significant. Although we show here the results for the ARFIMA processes only, the qualitatively similar results can be obtained for time series representing other examples of the uncorrelated statistical processes, which supports our claim that $\rho_q(s)$ correctly estimates the level of cross-correlations between uncorrelated signals.

\begin{figure}
\includegraphics[scale=0.48]{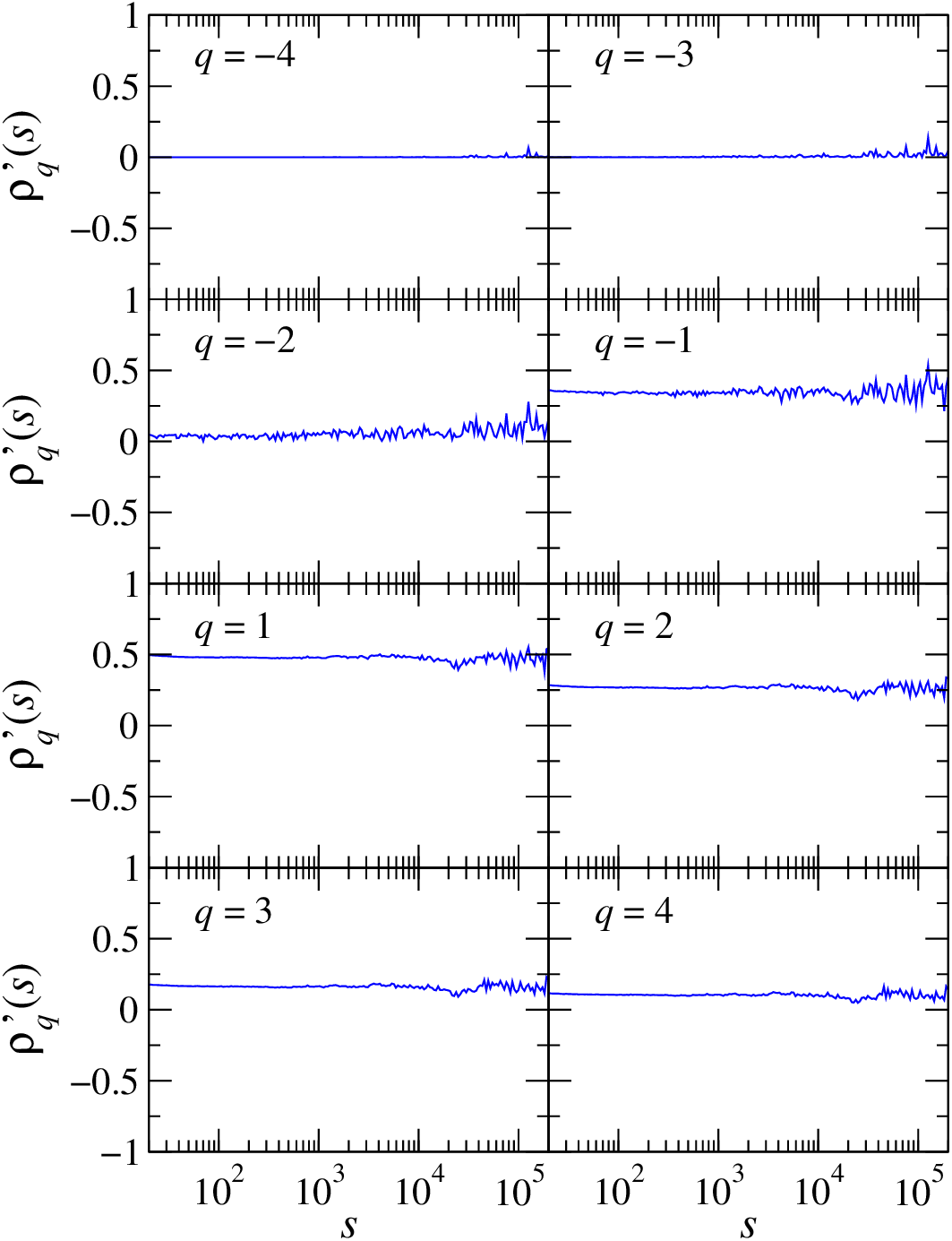}
\caption{(Color online) The incorrectly defined coefficient $\rho'_q(s)$ (using the no-sign definition of $F_{XY}^q(s)$ in Eq.~(\ref{eq::covariance.q})) calculated for the same pair of the randomized ARFIMA time series that was already used in Fig.~\ref{fig::arfima.shuffled}. Observe that for $q \ge -1$ the coefficient gives a false indication of positive cross-correlations. Apparent statistical significance of these results can be seen by comparing them with Fig.~\ref{fig::arfima.deviation}.}
\label{fig::arfima.shuffled.nosign}
\end{figure}

There is a good moment here to stress that a necessary condition for obtaining the correct results (i.e., the lack of cross-correlations) shown in Fig.~\ref{fig::arfima.shuffled} is a proper definition of the $q$-dependent fluctuation function $F_{XY}^q(s)$. Such a definition requires preserving the signs of the covariances $f_{XY}^2(s,\nu)$ exactly as it was done in Eq.~(\ref{eq::covariance.q}). Unfortunately, it often happens in literature that those signs are neglected and only the moduli of the covariances are considered (e.g.,~\cite{he2011}). What consequences can this approach have is illustrated in Fig.~\ref{fig::arfima.shuffled.nosign}. We take the same data as in Fig.~\ref{fig::arfima.shuffled} and calculate a modified coefficient $\rho'_q$ that uses the no-sign definition of $F_{XY}^q(s)$. As one can see, for $q \ge -1$ this coefficient falsely indicates the presence of statistically significant positive cross-correlations across all the scales, even though we know that such correlations cannot exist between the independently randomized signals.

\subsection{Partially cross-correlated signals}

The above examples of the almost perfectly correlated and the almost perfectly uncorrelated signals form the two extremes that are the least interesting trivial cases between which one can find a spectrum of much more interesting cases where the signals may reveal partial cross-correlations. For example, the correlations that are restricted to the signals' specific components or that are transient in time. We shall discuss here both these situations.

\begin{figure}
\includegraphics[scale=0.34]{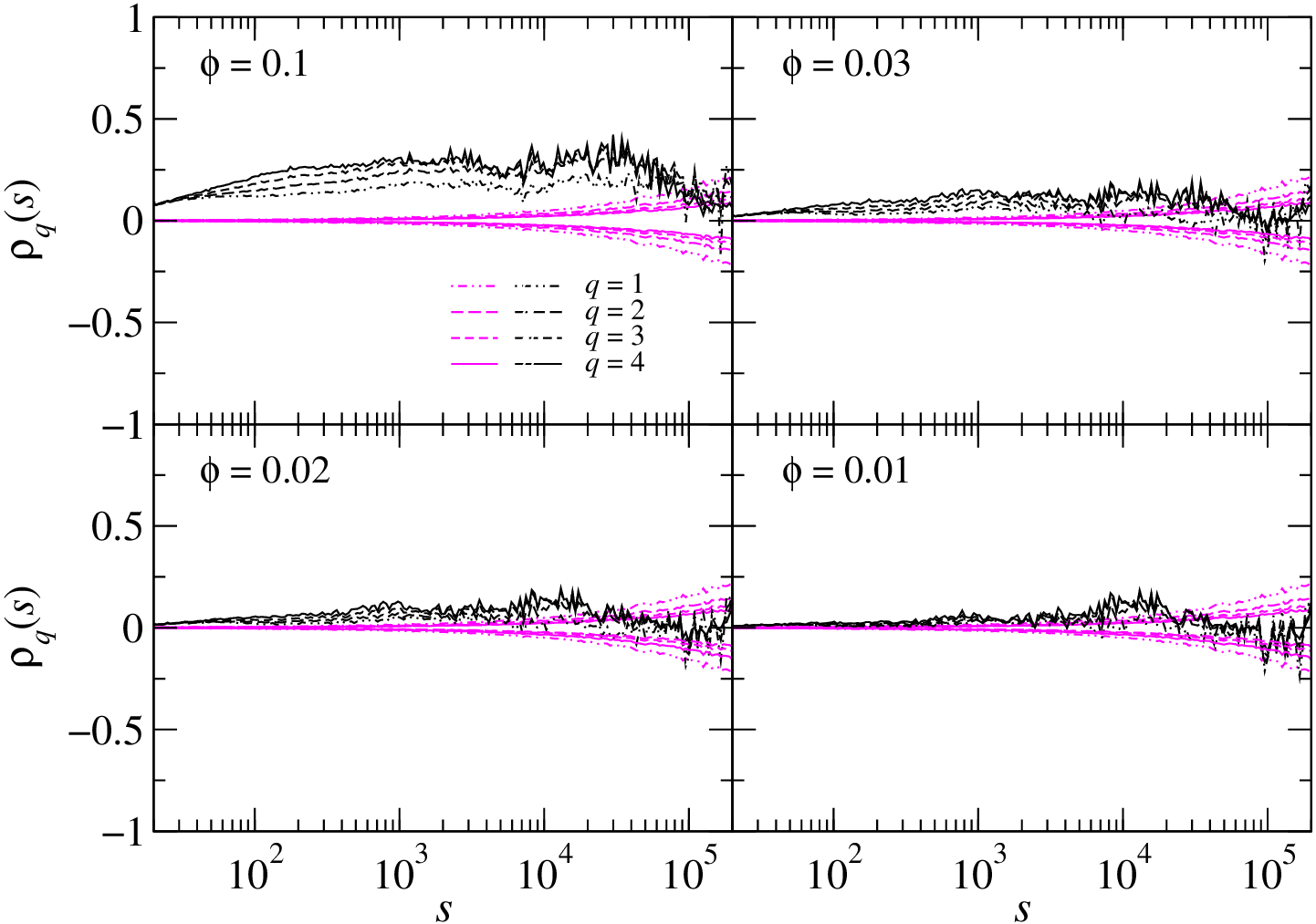}
\caption{(Color online) Sensitivity test for the $\rho_q(s)$ coefficient calculated based on the ARFIMA time series ($q > 0$). An original time series 1 is compared with a modified time series 2, in which only a fraction $\phi$ of the data points is preserved, while the rest is randomized (black lines). The results for the null-hypothesis of the uncorrelated time series are also shown for reference as the $+/-$ standard deviation of 100 independent realizations of the shuffled surrogates (magenta/grey lines symmetric with respect to $\rho_q=0$).}
\label{fig::sensitivity}
\end{figure}

First, we test sensitivity of the $q$DCCA approach if the cross-correlations are transient. In order to do this, we take two cross-correlated ARFIMA time series (with $\rho_q(s) \simeq 1$ as in Section IIIA) and randomize a part of the data points in one of these time series. After this operation, the modified time series consists of a fraction $\phi$ of the original data points in its center and a fraction $1-\phi$ of the shuffled points. Sample results of these calculations are shown in Fig.~\ref{fig::sensitivity} for different choices of $\phi$. It is clear that in this particular case the method gives statistically significant indication of the cross-correlated character of fluctuations for as low as $\phi=2\%$ correlated data points in both signals. This indication is restricted to the small and medium temporal scales ($s \le 10^3$) only, however. In order to obtain similar significance for the larger scales up to $s=10^5$, one has to increase the correlated fraction to $\phi > 0.1$, so that $s < \phi T$, where $T$ is the length of the time series.

Second, we assume that each time series under study consists of different components that have different amplitudes. Only some of these components are cross-correlated between the time series, while the other ones remain uncorrelated. For instance, a pair of time series may be cross-correlated only via large or via medium fluctuations and uncorrelated in small ones. We therefore expect that the $q$DCCA analysis will allow us: (i) to identify the existence of the cross-correlations and (ii) to show, what fluctuation amplitudes carry these cross-correlations.

The ARFIMA processes are not ideal for being a subject of the present analysis since their fluctuations are essentially Gaussian distributed and the range of their amplitudes is rather small. The method works fine also in this case, but here a much more instructive example can be the non-stationary processes with heavy-tailed fluctuations, where the amplitudes of the large and the small fluctuations differ considerably. The heavy-tailed fluctuations often appear in empirical data recorded from both natural and social systems, so our choice is realistic and it does not limit the applicability domain of the method.

Let us consider time series generated according to the Markov-switching multifractal model developed in~\cite{lux2006,liu2007}. This is a multiplicative, hierarchical model that is able to reproduce the multifractal characteristics of some types of empirical data (for example, the financial volatility). From the point of view of the present analysis, it is important that it produces the (unsigned) signals with heavy-tailed probability distribution functions, whose fluctuations can exhibit long-range, multiscale auto-correlations. According to the MSM model, an observable $x(i)$ is given by:
\begin{equation}
x(i) = \sigma(i) u(i),
\label{eq::msm}
\end{equation}
where $u(i)$ is a Gaussian random variable and $i$ plays a role of discrete time. $\sigma(i)$ is called the instantaneous volatility and defined as a product of a constant factor $\sigma$ and $k$ multipliers $M_j(i)$ drawn from the binary or the lognormal distribution:
\begin{equation}
\sigma^2(i) = \sigma^2 \prod_{j=1}^k M_j(i).
\end{equation}
In the binomial case, $M_j(i) = [m,2-m]$, $1 \le m \le 2$, while in the lognormal case: $M_j(i)= LN (-\lambda,2 \lambda)$~\cite{lux2006}. Each multiplier $M_j(i)$ changes its value at time $i$ with probability:
\begin{equation}
\gamma_j = 1 - (1 - \gamma_k)^{b^{j-k}}, \quad j=1,...,k,
\end{equation}
where $0 \le \gamma_k \le 1$ and $b = 2,3,4,...$ . Here we choose the cascades with $k=10$ levels and $b=2$ branches. We also omit the Gaussian-distributed factor $u(i)$ in Eq.~(\ref{eq::msm}) as being unnecessary from the point of view of our analysis, because the complexity of $x(i)$ is related to the instantaneous volatility $\sigma(i)$.

As a consequence of the stochastic character of the multiplier values, two independently produced signals would not be cross-correlated. A pair of the multiscale cross-correlated time series can nevertheless be obtained from an MSM time series by copying the set of its multipliers $M_j^{(1)}(i)$ and adding a small amount of noise to each one: $M_j^{(2)}(i) = M_j^{(1)}(i) + |\alpha \varepsilon(i)|$, where $\varepsilon(i)$ is an i.i.d. Gaussian random variable and $\alpha \ll 1$.

\begin{figure}
\includegraphics[scale=0.34]{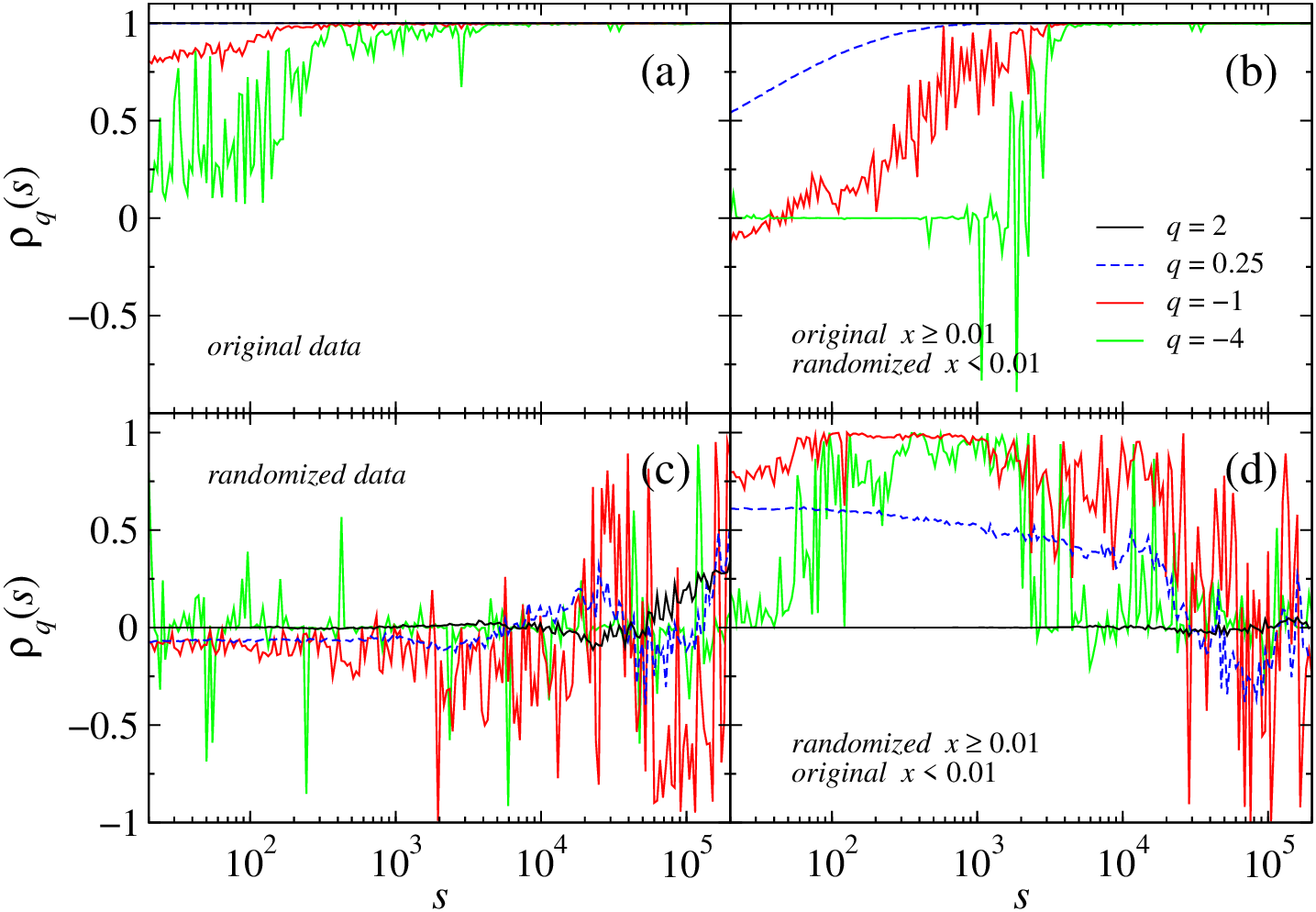}
\caption{(Color online) The $q$DCCA coefficient $\rho_q(s)$ calculated for a pair of time series from the lognormal Markov-switching multifractal (MSM) model with $k=10$, $b=2$, $\lambda=1.1$, and $\alpha=0.01$ (see text for the description of these parameters). Four selected values of $q$ are shown denoted by different lines. $q=2$ corresponds to the standard coefficient $\rho_{\rm DCCA}$. (a) The original time series. (b) Both time series are filtered (see text) in order to remove the correlations among the small fluctuations ($x(i) < 0.01$). (c) The randomized data (single realization of the shuffled surrogates) without any temporal correlations. (d) Both time series are filtered in order to remove the correlations among the medium and large fluctuations ($x(i) \ge 0.01$). For clarity of the pictures, the inverted values of $\rho_q(s)$ are not distinguished.}
\label{fig::msm.good.examples}
\end{figure}

Our objective is to show that the $q$DCCA coefficient $\rho_q(s)$ can provide us with more information on the cross-correlation structure of these time series that the ordinary coefficient $\rho_{\rm DCCA}(s)$ can. We generate two lognormal MSM time series of length $T=10^6$ with $\lambda=1.1$. The parameter $\alpha=0.01$ guarantees that their detrended fluctuations are cross-correlated. Indeed, Fig.~\ref{fig::msm.good.examples}(a) displays that for $q \in \langle -4,4 \rangle$ the coefficients $\rho_q(s)$ detect strong cross-correlations (for $q > 0$ even the maximum ones over all the scales). What then happens if we apply a filter that randomizes all the data points with small amplitude $x(i) < 0.01$? Action of this filter is equivalent to removing the cross-correlations among the small fluctuations and preserving the ones among the medium and large fluctuations. Fig.~\ref{fig::msm.good.examples}(b) presents the results obtained for such filtered signals. One can see that while the cross-correlations at the longer scales (roughly, for $s > 10^3$) are still detected by $\rho_q(s)$ for any $q$, for $q < 1$ the $q$DCCA coefficients indicate a significant decrease of the cross-correlations at the shorter scales ($s < 10^3$). This decrease is more explicit for $q < 0$ and for $q=-4$ the corresponding coefficient indicates that the signals are completely uncorrelated. As such small values of $q$ are related to very small fluctuations, this result is exactly what one might expect for the filtered signals. On the other hand, for $q \ge 2$ the filtering does not introduce any change in $\rho_q(s)$ and this is also perfectly compliant with the {\it a priori} expectation that for the large fluctuations the cross-correlations should survive. Now let us look at the results for $q=2$ in Fig.~\ref{fig::msm.good.examples}(a)(b): a change is hardly seen. Thus, in this case the $\rho_{\rm DCCA}$ coefficient proves totally insensitive to a significant change in the correlation structure of the data.

Another example of such an evident lack of sensitivity of $\rho_{\rm DCCA}$ can be seen in bottom panels of Fig.~\ref{fig::msm.good.examples}. If the same MSM time series as above are completely shuffled, all their temporal correlations are destroyed. Fig.~\ref{fig::msm.good.examples}(c) shows that the coefficients $\rho_q(s)$ do not detect any cross-correlations accordingly. The results for the large scales are within statistical error even if they are far from zero; among them the coefficient for $q=2$ is almost flat for the majority of scales except the largest ones. These results can be compared with the results obtained for the filtered signals (Fig.~\ref{fig::msm.good.examples}(d)), in which only the correlations among the small fluctuations ($x(i)<0.01$) are preserved and all the other correlations are removed via data shuffling. Now the coefficients $\rho_q(s)$ indicate the presence of cross-correlations for the medium scales (for $q=0.25$, $q=-1$, and $q=-4$) and for the small scales (for $q=0.25$ and $q=-1$). Such values of $q$ correspond to the small fluctuations. In contrast, for $q=2$ the related coefficient does not show anything significant and behaves roughly similar to the no-correlation case of Fig.~\ref{fig::msm.good.examples}(c).

\begin{figure}
\includegraphics[scale=0.34]{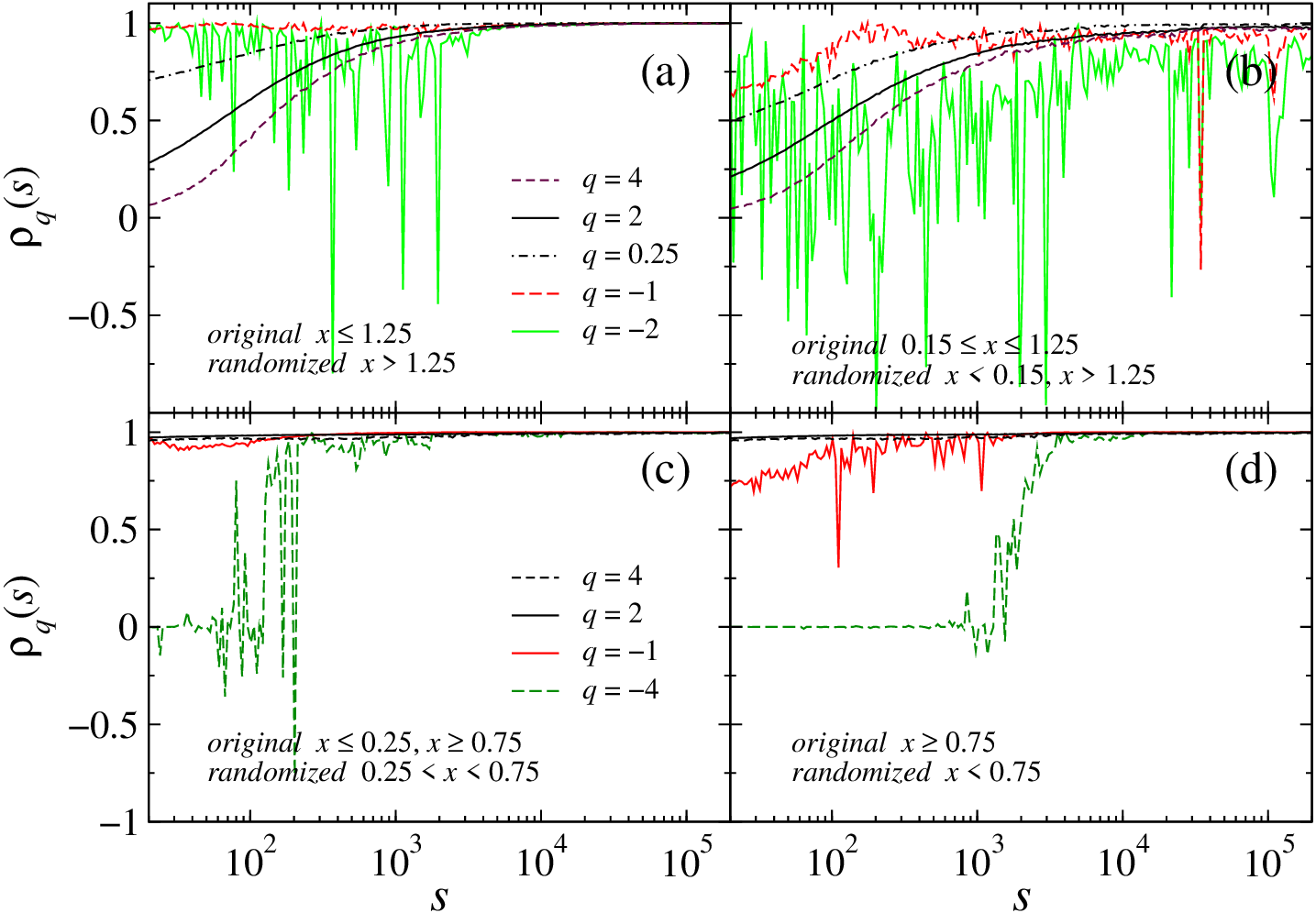}
\caption{(Color online) The $q$DCCA coefficient $\rho_q(s)$ calculated for a pair of time series from the binary Markov-switching multifractal (MSM) model with $k=20$ and $M_j(i)=[1.25,0.75]$ (cascade no. 1) or $M_j(i)=[1.2,0.8]$ (cascade no. 2). Different values of $q$ are denoted by different lines. The case of $q=2$ corresponds to the standard DCCA coefficient $\rho_{\rm DCCA}$. (Top) $\rho_q(s)$ for (a) a pair of time series with randomized fluctuations of large amplitudes $x>1.25$ and for (b) a pair of time series with randomized fluctuations of large amplitudes $x>1.25$ and small amplitudes $x<0.15$. (Bottom) $\rho_q(s)$ for (c) a pair of time series with randomized fluctuations of medium amplitudes $0.25 < x < 0.75$ and for (d) a pair of time series with randomized fluctuations of small and medium amplitudes $x<0.75$. For clarity of the pictures, the inverted values of $\rho_q(s)$ are not distinguished.}
\label{fig::msm.subtle.examples}
\end{figure}

The two further examples illustrating usefulness of the $q$DCCA coefficient are obtained from the binary MSM model producing a pair of the $k$=20-level binary multiplicative cascades with the $j$-level multipliers $M_j(i)$ drawn from the set $[1.2,0.8]$ and the set $[1.25,0.75]$, respectively. The resulting time series are of length $T=1,048,576$ and their values belong to the interval (0.01,100). In order to prepare the top panels of Fig.~\ref{fig::msm.subtle.examples}(a)(b), we modify both time series in such a way that we leave the original values if they do not exceed a threshold of $x=1.25$ and randomize the values above this threshold. Then we create copies of the so-modified time series and once again randomize their values if they are below $x=0.15$. By doing this, we get the two pairs of signals: (i) a pair of signals without the largest fluctuations, (ii) a pair of signals without the largest and the smallest fluctuations. Now we compare the coefficients $\rho_q(s)$ calculated for these pairs. First, we look at $\rho_2(s)$ (black solid lines). In both top panels, we see that $\rho_2(s)$ deviates significantly from unity for the scales $s < 10^3$, which suggests that the cross-correlations are not uniform across the fluctuation amplitudes. However, based solely on $q=2$ we cannot state decisively, which fluctuations correspond to this decreased correlations. As regards a comparison between the panels (a) and (b), there is some visible difference in $\rho_2(s)$ for the middle scales ($10^2 \le s \le 10^4$), but again one cannot decide what is the origin of this difference.

Let us now include $q \neq 2$. For $q=4$ the values of $\rho_q(s)$ in both panels are even smaller than their counterparts for $q=2$ and this is especially strong for the smallest scales $s < 10^2$. The opposite relation is seen for $q=0.25$, where $\rho_{0.25}(s)$ is much larger than $\rho_2(s)$. This means that by increasing $q$ we obtain a decrease of $\rho_q(s)$, which can correctly be interpreted as a manifestation of uncorrelated behavior of the largest fluctuations in both pairs of the signals. If we then compare the coefficients for $q=0.25$ and $q=-1$, we realize that they are systematically larger in (a) than in (b). Even for $q=-2$ the instability of $\rho_{-2}(s)$ suggest a lack of cross-correlations. All this means that the small fluctuations are weaker cross-correlated in (b) than in (a), exactly as it is expected from the signal construction in both cases. 

Next we once more use the time series generated by the binary MSM model. We randomize the fluctuations in the middle range of the amplitudes: $0.25 < x < 0.75$, which comprises the most frequent fluctuations in both time series, so that only the small and the large fluctuations remain cross-correlated. The results for this pair of signals are shown in Fig.~\ref{fig::msm.subtle.examples}(c). A copy of each time series is created and now the small fluctuations are a subject of additional randomization. Thus, we obtain the second pair of time series with only the large fluctuations that are still cross-correlated (Fig.~\ref{fig::msm.subtle.examples}(d)). The difference between these pairs is the existence (c) and the lack (d) of the cross-correlations among the fluctuations with small amplitude. On the one hand, for $q=2$ and $q=4$ we do not observe any difference in $\rho_q(s)$ between (c) and (d) as they are close to unity in both panels. On the other hand, the values of $\rho_q(s)$ for $q=-1$ and $q=-4$ are substantially smaller for small $s$ in (d) than in (c). Taking these results altogether, one can correctly conclude that both signal pairs are cross-correlated in the large fluctuations and uncorrelated in the small ones. It has to be stressed that drawing this conclusion would be impossible if we restricted our analysis to the $q=2$ case of $\rho_{\rm DCCA}$. (The uncorrelated character of the fluctuations of the medium amplitudes can be revealed in other comparative analyses, but here in this example such a range of amplitudes is beyond our interest.)

\section{Empirical data}
\label{sect.4}

\begin{figure}
\includegraphics[scale=0.43]{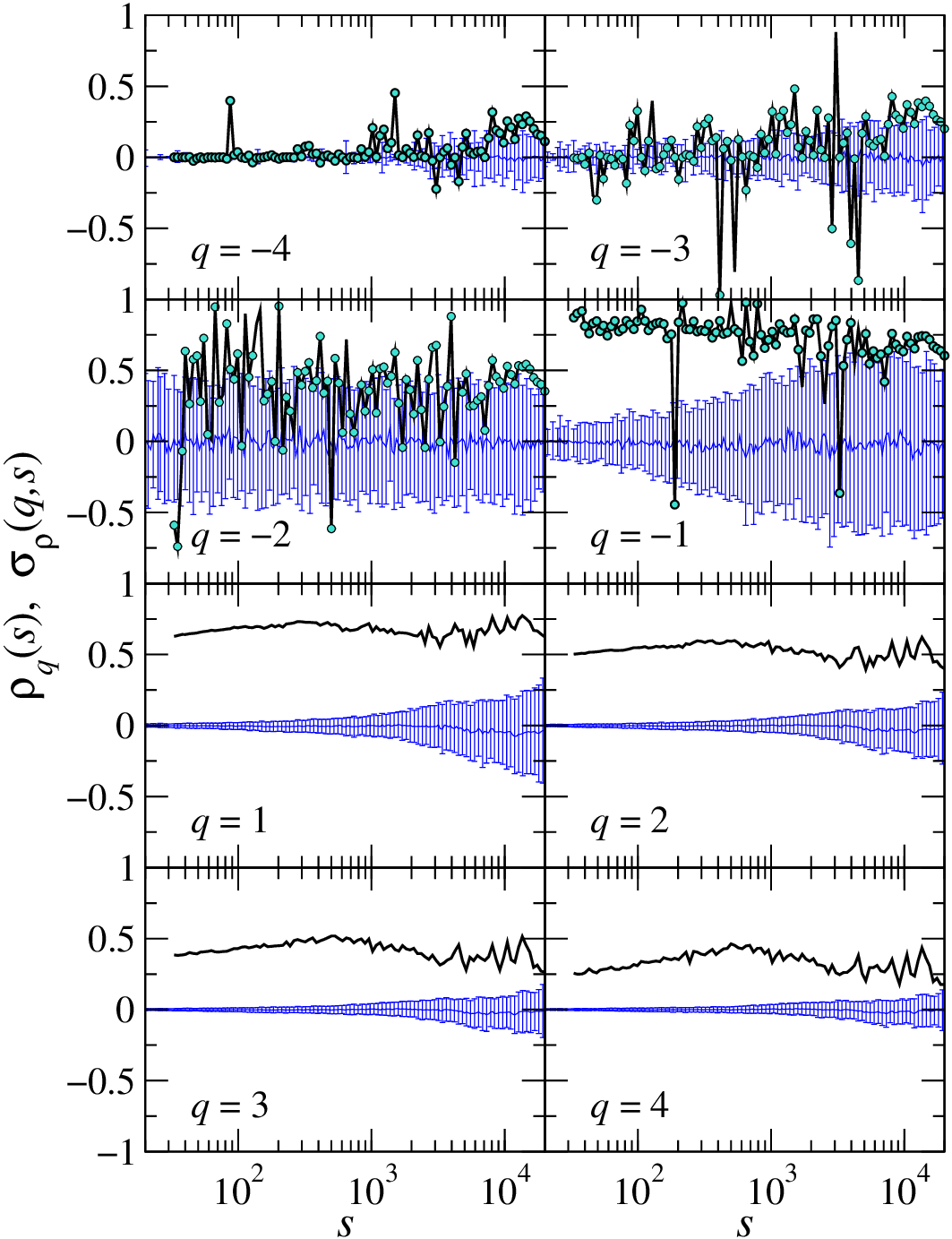}

\includegraphics[scale=0.43]{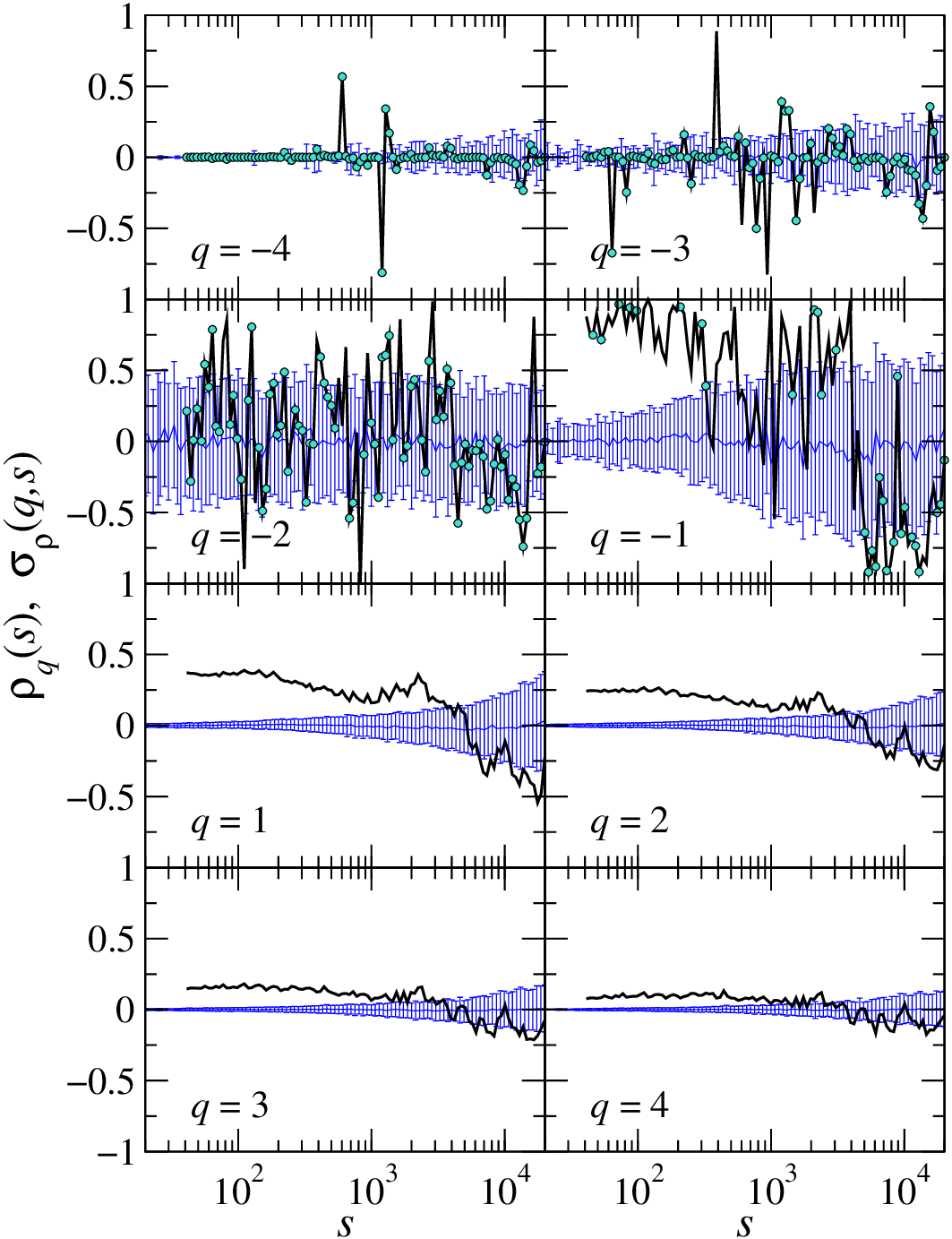}
\caption{(Color online) The $q$DCCA coefficient $\rho_q(s)$ calculated for two pairs of time series representing stock price logarithmic returns sampled with $\Delta t=5$ min frequency. (Top) Two stocks from the same industrial sector: Microsoft vs. Intel. (Bottom) Two stocks from different industrial sectors: Microsoft vs. 3M. Each panel for both pairs shows $\rho_q(s)$ calculated for different value of $q$ (heavy black line) together with the mean $\langle \rho_q(s) \rangle$ and the standard deviation $\sigma_{\rho}(q,s)$ obtained from 100 independent realizations of the shuffled surrogate data (thin blue/grey lines with error bars). The inverted values of $\rho_q(s)$ for $q < 0$ are denoted by circles.}
\label{fig::stocks}
\end{figure}

Financial data is known to be cross-correlated across different temporal scales~\cite{podobnik2008,zebende2011b,kristoufek2015b}, so it can serve as a suitable subject to show practical application of the $q$DCCA coefficient. We choose data from two large financial markets: the American stock market represented by the stocks traded at New York Stock Exchange (NYSE) or NASDAQ and the foreign currency exchange market (Forex), which is a global market. In the former case, our data set comprises high-frequency recordings of the stock prices corresponding to the 100 largest American companies over the years 1998-99. For each considered stock $i$, the corresponding time series represents the logarithmic price increments (returns) $r_i(t,\Delta t)= \log P_i(t+\Delta t) - \log P_i(t)$ sampled at a fixed time interval $\Delta t = 5$ min. The length of such time series is $T=40,638$ data points. The sample results for two pairs of stocks: (a) Microsoft$-$Intel and (b) Microsoft$-$3M are displayed in Fig.~\ref{fig::stocks} and tested against the null hypothesis that the cross-correlations are random (standard deviation of 100 independent realizations of the shuffled surrogates).

\begin{figure}
\includegraphics[scale=0.45]{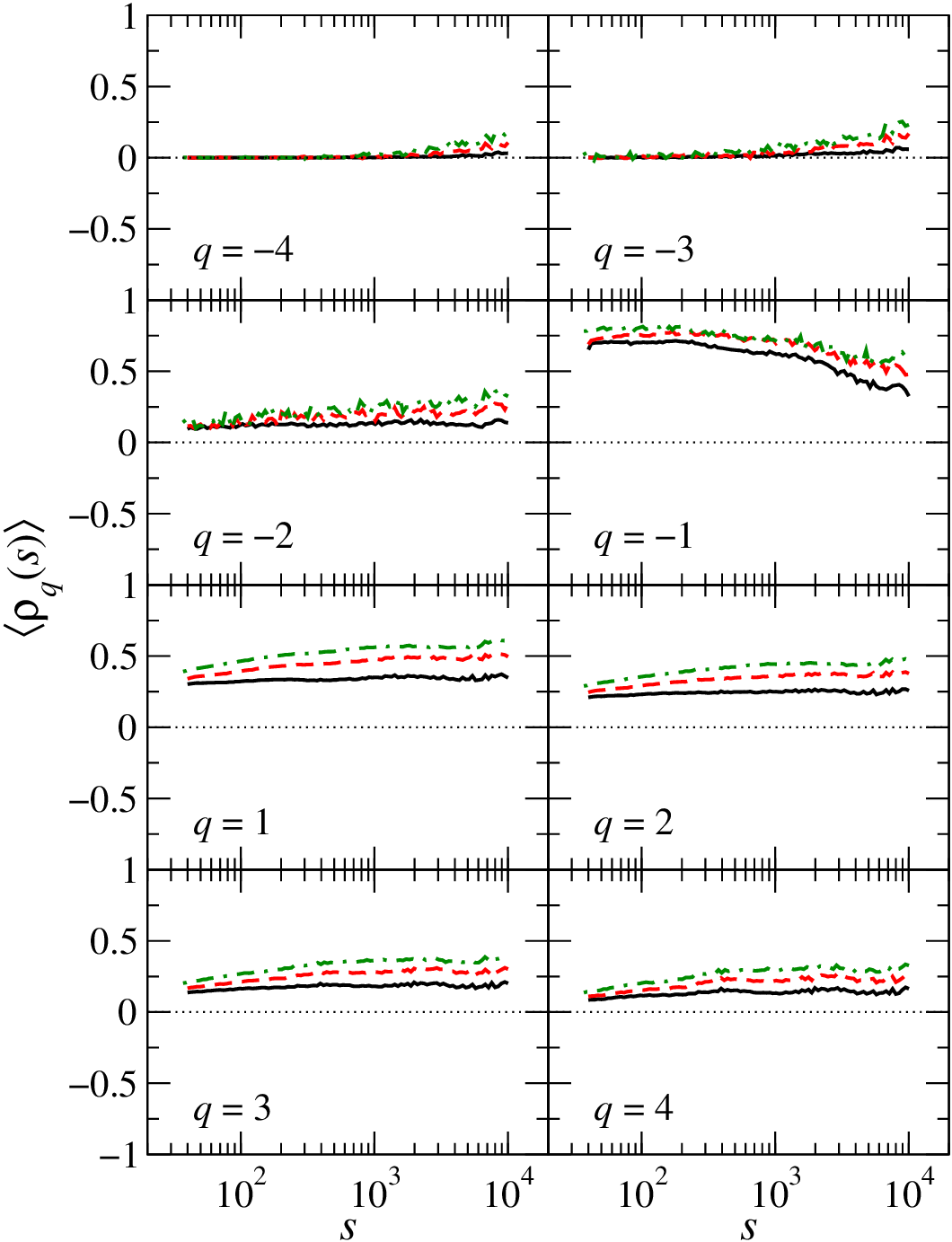}
\caption{(Color online) The average $q$DCCA coefficients $\langle \rho_q(s) \rangle$ calculated for the stock pairs divided into three groups: the stocks representing different industrial sectors (solid black), the stocks representing the same sector but different subsectors (dashed red/grey), and the stocks representing the same subsector (dash-dotted green/grey). Dotted lines denote the zero-correlation level.}
\label{fig::sectors}
\end{figure}

For $q \le -2$ corresponding to small fluctuations, there is no statistically significant indication of cross-correlations between any pair of stocks. Genuine cross-correlations are however visible starting from $q=-1$ up to high positive $q$s. One can see here that even if a vast majority of the $\rho_q(s)$ values are originally larger than 1 (see the red dots representing the inverted values of the coefficient for $q=-1$), the cross-correlations may be considered genuine as long as the function $\rho_q(s)$ is approximately stable (i.e., without strong fluctuations) over some range of the scales.

It is interesting to notice the weakening of the cross-correlations if one goes from small positive $q$s towards larger ones. It may be interpreted as a manifestation of the fact that the most cross-correlated fluctuations are those of the medium amplitudes, while the largest fluctuations are less correlated among the stocks. One can also infer that in both examples, the coefficients $\rho_q(s)$ are larger for the small scales and smaller for the large scales. However, it can be the opposite for different choice of the stocks (not shown) so there is no regularity here. By looking at both parts of Fig.~\ref{fig::stocks}, one can see that there is an important difference between them: the correlations are much stronger (larger values of $\rho_q(s)$) for the stocks representing mutually related industrial sectors (Microsoft$-$Intel, Fig.~\ref{fig::stocks}(a)), while they are weaker for the stocks from unrelated sectors (Microsoft$-$3M Fig.~\ref{fig::stocks}(b)).

That this is a systematic effect one can learn from Fig.~\ref{fig::sectors}, where three groups of the stock pairs are considered, representing different levels of the industrial similarity between companies: (i) the stocks corresponding to different industrial sectors, (ii) the stocks representing the same sector but different subsectors, and (iii) the stocks corresponding to the same subsector (according to~\cite{yahoo}). Obviously, the similarity level increases here gradually from (i) to (iii). We calculate $\rho_q(s)$ for each pair of stocks (4950 pairs total: 4218 pairs in (i), 548 pairs in (ii), and 184 pairs in (iii)) and then average the obtained functions over the pairs within each group. Fig.~\ref{fig::sectors} confirms that the values of $\langle \rho_q(s) \rangle$ systematically increase if the pairs consist of the stocks with increasing industrial similarity.

\begin{figure}
\includegraphics[scale=0.45]{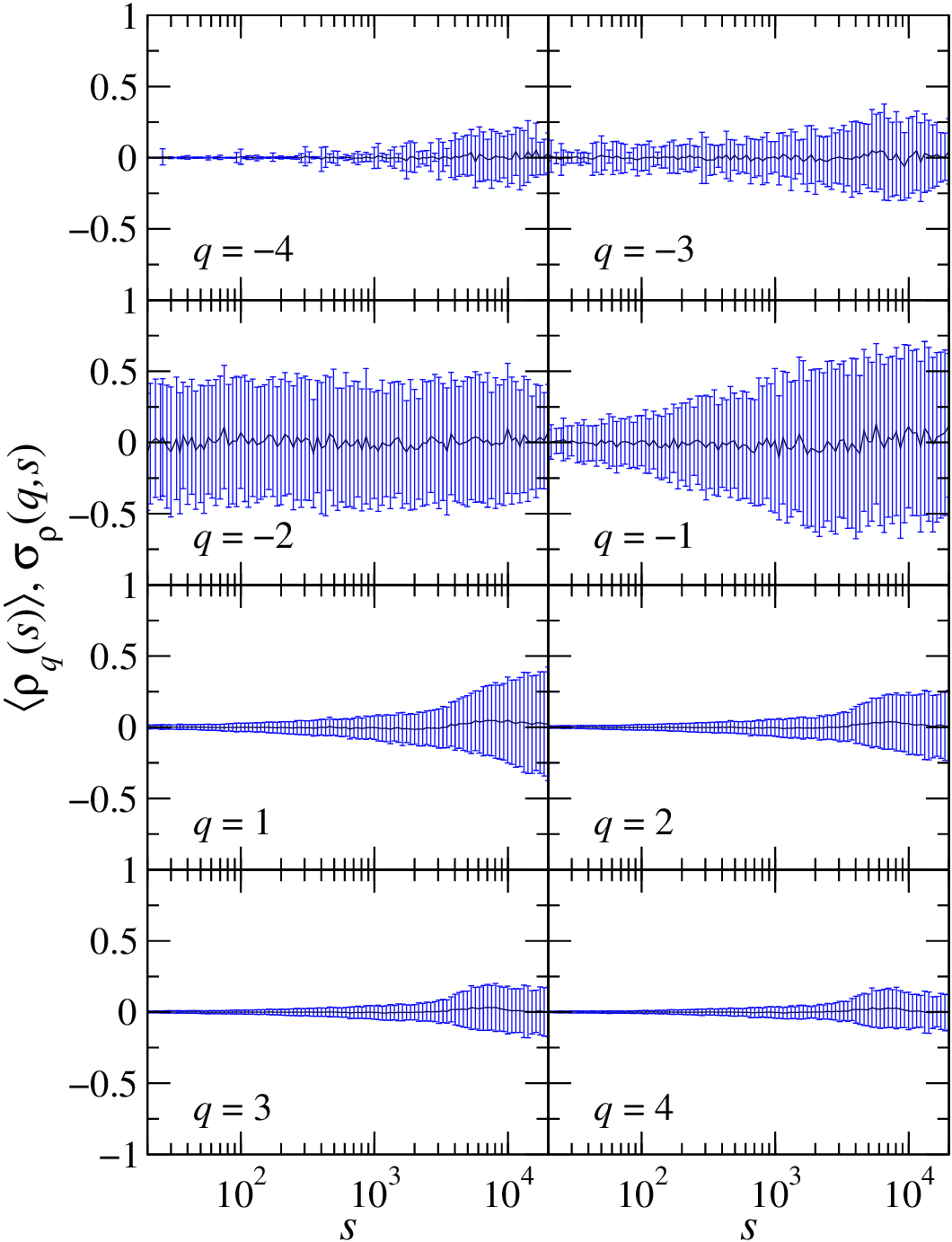}
\caption{(Color online) The average $q$DCCA coefficient $\langle \rho_q(s) \rangle$ and its standard deviation $\sigma_{\rho}(q,s)$ calculated for the Microsoft$-$Intel pair (the same as in Fig.~\ref{fig::stocks}(top)) after randomizing the Fourier phases of the original signals, which destroyed all the statistical dependences except for the Pearson autocorrelation function. 100 independent realizations of such surrogate data are used (thin blue/grey lines with error bars).}
\label{fig::fourier.surrogates}
\end{figure}

It is worth to mention that the correlations that lead to the non-zero values of $\rho_q(s)$ are nonlinear. We infer this from a failed test of the null hypothesis that the surrogate signals with the same Pearson autocorrelation function can reproduce the $\rho_q(s)$ functions. Fig.~\ref{fig::fourier.surrogates} documents the results of this test, showing that, on average, the Fourier-phase shuffling leads to complete destruction of the correlations that $\rho_q$ is sensitive to (compare with Fig.~\ref{fig::stocks}(a)). This is especially evident for the positive $q$s, for which the standard deviation $\sigma_{\rho}(q,s)$ obtained from 100 independent realizations of the surrogates is relatively small, especially for the small and medium scales $s$. For the large scales the MFDFA and MFCCA procedures are less efficient in detrending (the related polynomial degree $m=2$ is too low in this case) and this produces the non-zero deviations of $\langle \rho_q(s) \rangle$ observed in Fig.~\ref{fig::fourier.surrogates}.

\begin{figure}
\includegraphics[scale=0.45]{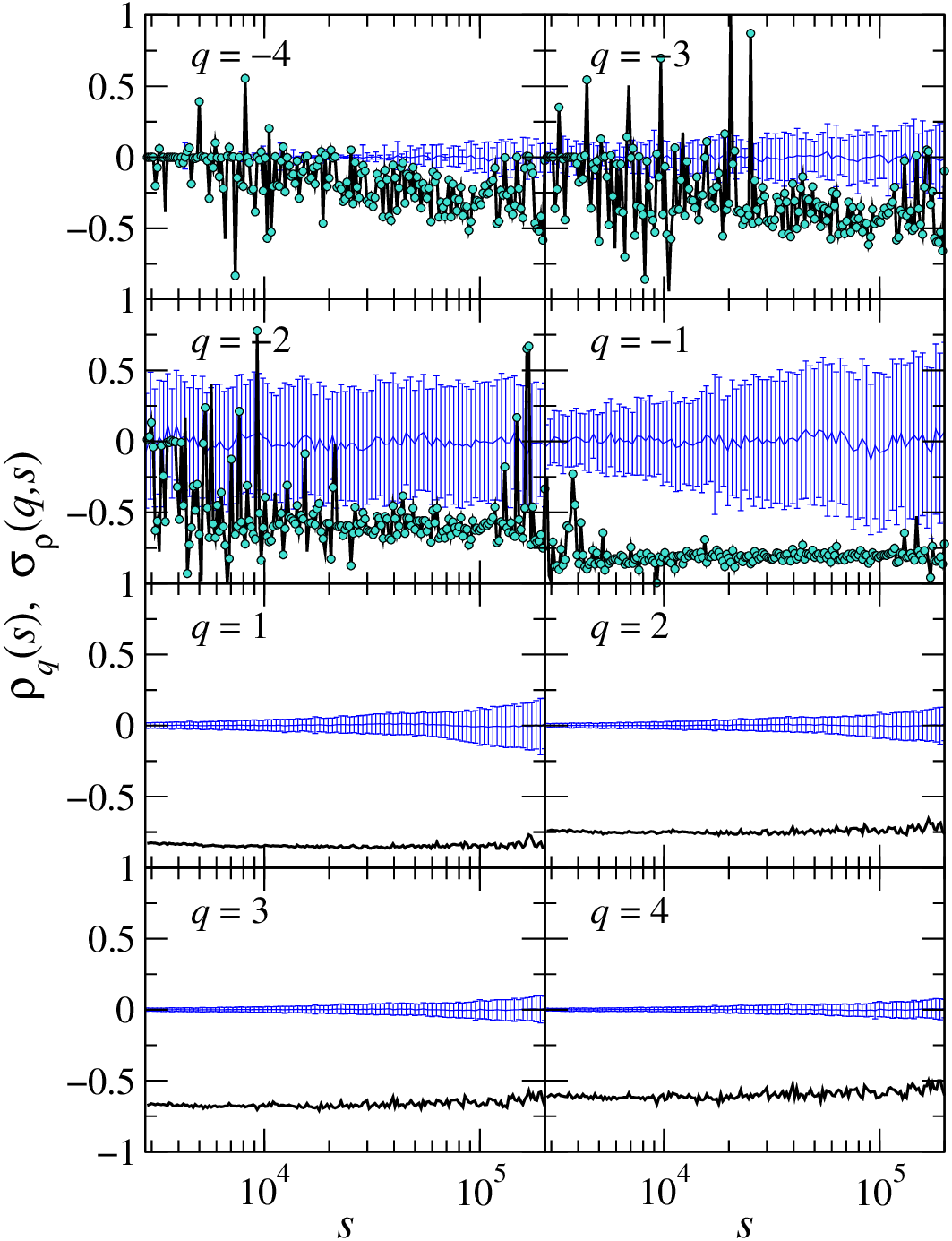}
\caption{(Color online) The $q$DCCA coefficient $\rho_q(s)$ calculated for a pair of time series representing returns of the USD/EUR and GBP/USD exchange rates sampled with $\Delta t=1$ min frequency. Each panel shows $\rho_q(s)$ calculated for different value of (heavy black line) together with the mean $\langle \rho_q(s) \rangle$ and the standard deviation $\sigma_{\rho}(q,s)$ obtained from 100 independent realizations of the shuffled surrogate data (thin blue/grey lines with error bars). The inverted values of $\rho_q(s)$ for $q < 0$ are denoted by circles.}
\label{fig::currencies}
\end{figure}

The second empirical example comes from the Forex. We consider high-frequency ($\Delta t$=1 min) logarithmic returns of the exchange rates among a set of major currencies recorded over the years 2004-2008. Fig.~\ref{fig::currencies} presents $\rho_q(s)$ calculated for two such time series corresponding to the USD/EUR and GBP/USD rates. Since euro and the British pound are positively cross-correlated (both are currencies used in the European Union countries) and since the US dollar appears in the numerator of the first rate and in the denominator of the second rate, these time series reveal negative cross-correlations of their detrended fluctuations. They are the strongest for $q=-1$ and $q=1$, while their strength decreases if we move away from these values in both directions. One can also see in the topmost panels of Fig.~\ref{fig::currencies} that even for $q < -2$ there is statistically significant indication of genuine cross-correlations for large $s$ (if compared with the standard deviation of the shuffled surrogates). We infer from these results that the most strongly cross-correlated fluctuations of the Forex data are those of the medium amplitudes. The large-amplitude fluctuations are relatively weaker but still substantially cross-correlated and even the fluctuations with the small amplitudes reveal weak cross-correlations (unlike their uncorrelated counterparts from the stock market in Fig.~\ref{fig::stocks}).

\section{Summary}
\label{sect.5}

We proposed a new measure of the multiscale detrended cross-correlations between a pair of time series, called the $q$DCCA coefficient. This coefficient forms actually a family of functions $\rho_q(s)$ that depend on the exponent $q$ and the temporal scale $s$. Like in the case of the multifractal, $q$-dependent fluctuation functions, by varying a value of $q$, we can amplify those fluctuations that are of specific amplitudes: e.g., the small ones for $q \ll 0$ or the large ones for $q \gg 0$, and study the structure of the cross-correlations among such fluctuations. A specific case of $\rho_q$ for $q=2$ corresponds to the DCCA coefficient $\rho_{\rm DCCA}$ already known from literature~\cite{zebende2011b}, so from this perspective the $\rho_q$ coefficient may be viewed as its generalization. For $q \ge 0$, the interpretation of $\rho_q$ is straightforward as its value is confined in the range $-1 \le \rho_q \le 1$: the maximum value of $\rho_q=1$ means that the corresponding detrended fluctuations of both signals under study show exactly the same correlation structure, $\rho_q=-1$ means that such fluctuations are completely anticorrelated, and $\rho_q=0$ indicates that they are independent of each other. For $q < 0$, on the other hand, the coefficient $|\rho_q|$ may assume arbitrarily large values, which can lead to interpretation problems, but this may happen especially if the detrended fluctuations of the two signals are uncorrelated. Knowing that, we may be concerned about the exact values of $\rho_q$ in this case. For practical reasons, we overcome this problem by replacing the original values $|\rho_q|>1$ by their inversion and thus forcing them to fit into the normal interval $\langle -1,1 \rangle$. In the case of no correlations, a course of the function $\rho_q(s)$ becomes strongly unstable and its values wildly fluctuate. This effect can thus serve as an optical indication of the signal independence. In contrast, if the signal fluctuations are cross-correlated, we obtain largely stable behavior of $\rho_q(s)$ even for $q<0$ and even if the original values of $|\rho_q(s)|$ slightly exceed 1. Therefore, in this case a lack of wild fluctuations of $\rho_q(s)$ and its non-zero value can signalize the existence of cross-correlations. All this means that the family of the coefficients $\rho_q$ is a well-defined tool of the multiscale detrended cross-correlation analysis. Obviously, because of the more demanding interpretation of results for $q < 0$, one may restrict an analysis to the $q \ge 0$ case, but this is equivalent to ignoring the cross-correlation structure of small fluctuations, so we do not recommend it.

In order to illustrate the performance of this new measure, we applied it to a few selected data sets representing long-memory processes: ARFIMA and the Markov-switching multifractal. We showed that $\rho_q$ is able to identify correctly specific cross-correlations that remain undetected by the standard DCCA cross-correlation coefficient $\rho_{\rm DCCA}$ as well as it is able to distinguish between the processes whose detrended fluctuations are cross-correlated in different amplitude ranges $-$ a task that is infeasible with $\rho_{\rm DCCA}$. We also carried out the $\rho_q$ analysis based on sample empirical data from the financial markets and found that the detrended cross-correlations among such data also depend on the detrended-fluctuation amplitude. Straightforward and likely fruitful for a more global correlation analysis would be to form the $q$-dependent counterparts of the conventional correlation matrices for larger sets of multivariate data and to study their spectral properties as it is commonplace in the standard correlation matrix analysis. By proceeding further in this direction, one may then also consider constructing the $q$-dependent graphs. On the other hand, a parallel approach would be defining the $q$-generalized version of the DMCA coefficient $\rho_{\rm DMCA}$~\cite{kristoufek2014b} in a full analogy to our approach.


\begin{thebibliography}{99}

\bibitem{mikosch2004} T.~Mikosch, C.~St\u aric\u a, Rev. Econ. Stat. {\bf 86}, 378 (2004).
\bibitem{bassler2007} K.E. Bassler, J.L. McCauley, G.H. Gunaratne, Proc. Natl. Acad. Sci. USA {\bf 104}, 17287 (2007).
\bibitem{pagan2002} A.R. Pagan, J. Econometrics {\bf 45}, 267 (2002).
\bibitem{schmitt2013} T.A. Schmitt, D. Chetalova, R.Sch\"afer, T. Guhr, EPL {\bf 103}, 58003 (2013).
\bibitem{mayer-kress1994} G. Mayer-Kress, Integ. Physiol. Behav. Sci. {\bf 29}, 205-210 (1994).
\bibitem{peng1995} C.‐K. Peng, S. Havlin, H.E. Stanley, A.L. Goldberger, Chaos {\bf 5}, 82 (1995).
\bibitem{kohlmorgen2000} J. Kohlmorgen, K.-R. M\"uller, J. Rittweger, K. Pawelzik, Biol. Cybern. {\bf 83} 73 (2000).
\bibitem{bernaola-galvan2001} P. Bernaola-Galv\'an, P.Ch. Ivanov, L.A.N. Amaral, H.E. Stanley, Phys. Rev. Lett. {\bf 87}, 168105 (2001).
\bibitem{perello2008} J.~Perell\'o, J.~Masoliver, A.~Kasprzak, R.~Kutner, Phys.~Rev.~E {\bf 78} 036108 (2008).
\bibitem{kwapien2014} J.~Kwapie\'n, S.~Dro\.zd\.z, Phys. Rep. {\bf 515}, 115 (2012).
\bibitem{peng1994} C.-K. Peng, S.V. Buldyrev, S. Havlin, M. Simons, H.E. Stanley, A.L. Goldberger, Phys. Rev. E {\bf 49}, 1685 (1994).
\bibitem{podobnik2008} B. Podobnik, H.E. Stanley, Phys. Rev. Lett. {\bf 100}, 084102 (2008).
\bibitem{podobnik2009} B. Podobnik, D. Horvatic, A.M. Petersen, H.E. Stanley, Proc. Natl. Acad. Sci. USA {\bf 106}, 22079 (2009).
\bibitem{xu2010} N. Xu, P. Shang, S. Kamae, Nonlin. Dyn. {\bf 61}, 207 (2010).
\bibitem{zebende2011a} G.F. Zebende, P.A. da Silva, A. Machado Filhoa, Physica A {\bf 390}, 1677 (2011).
\bibitem{qian2015} X.-Y. Qian, Y.-M. Liu, Z.-Q. Jiang, B. Podobnik, W.-X. Zhou, H.E. Stanley, arXiv:1504.02435 (2015).
\bibitem{kantelhardt2002} J.W. Kantelhardt, S.A. Zschiegner, A. Bunde, S. Havlin, E. Koscielny-Bunde, H.E. Stanley, Phys. A {\bf 316}, 87 (2002).
\bibitem{oswiecimka2005} P.~O\'swi\c ecimka, J.~Kwapie\'n, S.~Dro\.zd\.z, Physica A {\bf 347}, 626 (2005).
\bibitem{muzy2008} J.F.~Muzy, E.~Bacry, R.~Baile, P.~Poggi, Europhys. Lett. {\bf 82}, 60007 (2008).
\bibitem{ivanov1999} P.Ch.~Ivanov, L.A.N.~Amaral, A.L.~Goldberger, S.~Havlin, M.G.~Rosenblum, Z.R.~Struzik, H.E.~Stanley, Nature {\bf 399}, 461 (1999).
\bibitem{udovichenko2002} V.V.~Udovichenko, P.E.~Strizhak, Theor. Exp. Chem. {\bf 38}, 259 (2002).
\bibitem{witt2013} A. Witt and B. D. Malamud, Surv. Geophys. 34, 541 (2013).
\bibitem{calvet2002} L.~Calvet, A.~Fisher, Rev. Econ. Stat. {\bf 84}, 381 (2002).
\bibitem{drozdz2009} S.~Dro\.zd\.z, J.~Kwapie\'n, P.~O\'swi\c ecimka, R.~Rak, EPL {\bf 88}, 60003 (2009).
\bibitem{drozdz2010} S.~Dro\.zd\.z, J.~Kwapie\'n, P.O\'swi\c ecimka, R.~Rak, New~J.~Phys. {\bf 12}, 105003 (2010).
\bibitem{koscielny2006} E.~Koscielny-Bunde, J.W.~Kantelhardt, P.~Braund, A.~Bunde, S.~Havlin, J.~Hydrology {\bf 322}, 120 (2006).
\bibitem{kantelhardt2006} J.W.~Kantelhardt, E.~Koscielny-Bunde, D.~Rybski, P.~Braun, A.~Bunde, S.~Havlin, J.~Geophys. Res. Atmos. {\bf 111}, D01106 (2006).
\bibitem{ausloos2012} M.~Ausloos, Phys. Rev.~E {\bf 86}, 031108 (2012).
\bibitem{jafari2007} G.R.~Jafari, P.~Pedram, L.~Hedayatifar, J.~Stat. Mech. P04012 (2007).
\bibitem{oswiecimka2011} P.~O\'swi\c ecimka, J.~Kwapie\'n, I.~Celi\'nska, S.~Dro\.zd\.z, R.~Rak, arXiv:1106.2902 (2011).
\bibitem{grech2008} D.~Grech, G.~Pamu{\l}a, Physica A {\bf 387}, 4299 (2008).
\bibitem{maiorino2015} E.~Maiorino, L.~Livi, A.~Giuliani, A.~Sadeghian, A.~Rizzi, Physica A {\bf 428}, 302 (2015).
\bibitem{makowiec2009} D.~Makowiec, A.~Dudkowska, R.~Ga{\l}\c aska, A.~Rynkiewicz, Physica A {\bf 388}, 3486 (2009).
\bibitem{kristoufek2015a} L.~Kristoufek, Phys.~Rev.~E {\bf 91}, 022802 (2015).
\bibitem{drozdz2015} S.~Dro\.zd\.z, P.~O\'swi\c ecimka, Phys.~Rev.~E {\bf 91}, 030902(R) (2015).
\bibitem{rotundo2015} G.~Rotundo, M.~Ausloos, C.~Herteliu, B.~Ileanu, Physica A {\bf 429}, 109 (2015).
\bibitem{oswiecimka2006} P.~O\'swi\c ecimka, J.~Kwapie\'n, S.~Dro\.zd\.z, Phys.~Rev.~E {\bf 74}, 016103 (2006).
\bibitem{zhou2008} W.-X.~Zhou, Phys. Rev.~E {\bf 77}, 066211 (2008).
\bibitem{jiang2011} Z.-Q.~Jiang, W.-X.~Zhou, Phys. Rev.~E {\bf 84}, 016106 (2011).
\bibitem{he2011} L.-Y.~He, S.-P.~Chen, Physica A {\bf 390}, 297 (2011).
\bibitem{li2012} Z.~Li, X.~Lu, Physica A {\bf 391}, 3930 (2012).
\bibitem{wang2012} G.-J.~Wang, C.~Xie, Acta Phys. Pol.~B {\bf 43}, 2021 (2012).
\bibitem{oswiecimka2014} P.~O\'swi\c ecimka, S.~Dro\.zd\.z, M.~Forczek, S.~Jadach, J.~Kwapie\'n, Phys. Rev.~E {\bf 89}, 023305 (2014).
\bibitem{chen-hua2015} S.~Chen-hua, L.~Chao-ling, S.~Ya-li, Physica A {\bf 419}, 417 (2015).
\bibitem{zebende2011b} G.F.~Zebende, Phys.~A {\bf 390}, 614 (2011).
\bibitem{vassoler2012} R.T.~Vassoler, G.F.~Zebende, Phys.~A {\bf 391}, 2438 (2012).
\bibitem{zebende2013} G.F.~Zebende, M.F.~da Silva, A.~Machado Filho, Phys.~A {\bf 392}, 1756 (2013).
\bibitem{reboredo2014} J.C. Reboredo, M.A. Rivera-Castro, G.F. Zebende, Energy Econ. {\bf 42}, 132 (2014).
\bibitem{kristoufek2014a} L.~Kristoufek, Physica A {\bf 402}, 291-298 (2014).
\bibitem{kristoufek2014b} L.~Kristoufek, Physica A {\bf 406}, 169-175 (2014).
\bibitem{podobnik2011} B.~Podobnik, Z.-Q.~Jiang, W.-X.~Zhou, H.E.~Stanley, Phys. Rev.~E {\bf 84}, 066118 (2011).
\bibitem{hosking1981} J.~Hosking, Biometrika {\bf 68}, 165 (1981).
\bibitem{gil-alana2012} L.A.~Gil-Alana, J.~Appl. Meteor. Climatol. {\bf 51}, 1904 (2012).
\bibitem{yusof2015} F.~Yusof, I.L.~Kane, Z.~Yusop, AIP Conf. Proc. {\bf 1643}, 446 (2015).
\bibitem{burnecki2014} K.~Burnecki, A.~Weron, J.~Stat. Mech. {\bf 2014}, P10036 (2014).
\bibitem{torre2007} K.~Torre, D.~Deligni\`eres, L.~Lemoine, Brit. J. Math. Stat. Psychol. {\bf 60}, 85 (2007).
\bibitem{leite2013} A.~Leite, A.P.~Rocha, M.E.~Silva, Chaos {\bf 23}, 023103 (2013).
\bibitem{saakian2012} D.B.~Saakian, Phys.~Rev.~E {\bf 85}, 031142 (2012).
\bibitem{lux2006} T.~Lux, {\it The Markov-switching multifractal model of asset returns: GMM estimation and linear forecasting of volatility}, Economic working paper No. 2006/17 (Christian-Albrechts-Universit\"at Kiel, 2006).
\bibitem{rypdal2011} B.~Rypdal, K.~Rypdal, J.~Goephys. Res. {\bf 116}, A02202 (2011).
\bibitem{calvet2008} L.~Calvet, A.~Fisher, {\it Multifractal volatility: Theory, forecasting, and pricing}, Academic Press (2008).
\bibitem{liu2007} R.~Liu, T.~Di Matteo, T.~Lux, Physica A {\bf 383}, 35 (2007).
\bibitem{yahoo} Yahoo! Finance, {\it http://finance.yahoo.com} .
\bibitem{kristoufek2015b} L.~Kristoufek, Physica A {\bf 428}, 194 (2015).

\end{thebibliography}
\end{document}